\documentclass[useAMS,usenatbib]{mn2e}
\usepackage{aas_macros}
\usepackage{mathrsfs}
\usepackage{graphicx}
\usepackage{amsmath}
\usepackage{amssymb}

\usepackage{times}
\usepackage[usenames,dvipsnames]{color}
\usepackage{ulem}

\begin{document}

\title[Prestellar Discs in Isolated and Multiple Prestellar Systems]
      {The Properties of Prestellar Discs in Isolated and Multiple Prestellar Systems}

\author[T.~Hayfield et al.]
{T.~Hayfield~$^{1,2}$\thanks{hayfield@mpia.de}, L.~Mayer~$^{3}$, J.~Wadsley~$^{4}$, and A.~C.~Boley~$^{5}$ \\
$^{1}$ Institute of Astronomy, Physics Department, ETH Z\"urich, Wolfgang-Pauli-Strasse 27, CH--8093 Z\"urich, Switzerland\\
$^{2}$ Max-Planck-Institut f\"ur Astronomie, K\"onigstuhl 17, D-69117 Heidelberg, Germany \\
$^{3}$  Institute of Theoretical Physics, University of Z\"urich,
  Winterthurerstrasse 190, CH--8057 Z\"urich, Switzerland \\ 
$^{4}$ Department of Physics \& Astronomy, McMaster University, 
  1280 Main St. West, Hamilton ON L8S 4M1 Canada\\
$^{5}$ Astronomy Department, University of Florida, 211 Bryant Space Science
  Center, PO Box 112055, USA
}

\date{Accepted . Received ; in original form .} 
\volume{000}
\pagerange{000--000} 
\pubyear{0000} 

\maketitle 

\def\msl#1{#1\rm\,M_\odot}
\def\mslyr#1{#1\rm\,M_\odot\,yr^{-1}}
\def\tffm{\sqrt{3\pi\!/32G\rho_{cl}}}

\def\jmm{{\rm\,cm^2\,s^{-1}}}
\def\gcc{{\rm\,g\,cc^{-1}}}
\def\tffs{{\rm t_{ff}}}
\def\rhoa{10^{-13}}
\def\rhob{5.7\times10^{-8}}
\def\rhoc{10^{-3}}

\def\ud{\mathrm{d}}
\def\deriv#1#2{\frac{\ud #1}{\ud #2}}

\def\mclf{\msl{50}}
\def\rclf{0.188{\rm\,pc}}
\def\sigclf{1.17{\rm\,km\,s^{-1}}}
\def\tfff{1.92\times10^{5}{\rm\,yr}}
\def\tstir{0.88\tffs}
\def\tbirth{\tstir}
\def\tbinff{0.893\tffs}

\def\tbinan{3330{\rm\,yr}}

\def\rhofill{7\times10^{-15}\gcc}
\def\rhofilh{10^{-13}\gcc}

\def\mfrag{\msl{8\times10^{-3}}}
\def\mjfrag{\msl{6.4\times10^{-3}}}

\def\tsim{1.2\,\tffs}
\def\tcl{10{\rm\,K}}
\def\rhocl{1.2\times 10^{-19}\gcc}
\def\rhothresh{6.7\times 10^{-12}\gcc}
\def\ebind{2.6\times 10^{34}{\rm\,erg}}
\def\soft{2.3{\rm\,AU}}
\def\taufac{7.3}

\def\tfi{1.7{\rm\,kyr}}
\def\tpa{4.4{\rm\,kyr}}
\def\tpb{5.8{\rm\,kyr}}
\def\tpc{7.1{\rm\,kyr}}
\def\tpd{9.2{\rm\,kyr}}
\def\tpe{10{\rm\,kyr}}
\def\tpf{11{\rm\,kyr}}
  
\def\pptrop{ P = \begin{cases}
    \kappa_0\rho       &  \rho < \rhoa, \\
    \kappa_1\rho^{7/5} &  \rhoa \leq \rho < \rhob, \\
    \kappa_2\rho^{1.15}&  \rhob \leq \rho < \rhoc, \\
    \kappa_3\rho^{5/3} &  \rhoc \leq \rho,
  \end{cases} }

\def\pqtrop{ P = \begin{cases}
    \kappa_0\rho       &  \rho < \rhoa, \\
    \kappa_1\rho^{7/5} &  \rhoa \leq \rho,
  \end{cases} }

\def\csnaught{1.84 \times 10^4 {\rm\,cm\,s^{-1}}}

\def\mcla{5.5\rm\,M_{jup}}
\def\mclaa{39\rm\,M_{jup}}

\def\mclb{7.4\rm\,M_{jup}}
\def\mclbb{14\rm\,M_{jup}}

\def\fw{0.47}
\def\figdir{figures/}

\def\rname#1{run #1}
\def\rnames#1{runs #1}
\def\Rname#1{Run #1}
\def\Rnames#1{Runs #1}

\def\altmergerdt{6.3{\rm\,kyr}}
\def\alttcol{1.1{\rm\,t_{ff}}}

\def\altfillen{7400{\rm\,AU}}
\def\altfilfragdt{2{\rm\,kyr}}
\def\altbox{(10^4{\rm\,AU})^3}
\def\alttpa{0{\rm\,kyr}}
\def\alttpb{4.3{\rm\,kyr}}
\def\alttpc{8.7{\rm\,kyr}}
\def\altmcl{12\rm\,M_{jup}}
\def\altjclust{2 \times 10^{19}\jmm}
\begin{abstract}

We present high-resolution 3D smoothed particle hydrodynamics
simulations of the formation and evolution of protostellar discs in a
turbulent molecular cloud.  Using a piecewise polytropic equation of
state, we perform two sets of simulations.  In both cases we find that
isolated systems undergo a fundamentally different evolution than
members of binary or multiple systems. When formed, isolated systems
must accrete mass and increase their specific angular momentum,
leading to the formation of massive, extended discs, which undergo
strong gravitational instabilities and are susceptible to disc
fragmentation. Fragments with initial masses of $\mcla$, $\mclb$ and
$\altmcl$ are produced in our simulations. In binaries and small
clusters, we observe that due to competition for material from the
parent core, members do not accrete significant amounts of high
specific angular momentum gas relative to isolated systems.  We find
that discs in multiple systems are strongly self-gravitating but that
they are stable against fragmentation due to disc truncation and mass
profile steeping by tides, accretion of high specific angular momentum
gas by other members, and angular momentum being redirected into
members' orbits. In general, we expect disc fragmentation to be less
likely in clusters and to be more a feature of isolated systems.
\end{abstract}

\section{Introduction}
The overall paradigm under which star formation occurs remains under debate. One
hypothesis is that stars are thought to form as a collection of low mass
fragments in collapsing clumps of gas, and then undergo {\em competitive accretion} 
as the fragments try to accrete gas from their common reservoir 
(\citealt{bonnellbateclarke01}; \citealt{bonnellbate06}). The main competing
hypothesis is {\em gravitational collapse}, where massive star-forming clumps
collapse and form multiple cores. Each star forms from the gas that is available
in its own core, with limited accretion from other material in the parent clump
\citep{krumholzmckeeklein05}.  Regardless of the star formation paradigm, in the
case of solar-type stars, the end result is most likely membership in a binary
or multiple system (\citealt{duquennoymayor91};
\citealt{eggenbergerudrymayor04}). Scenarios for the formation of such systems
abound. Amongst the simplest, most idealised scenarios are the fission of a bar-unstable core
(\citealt{durisengingoldtohlineboss86};
\citealt{burkertbatebodenheimer97}) and the fragmentation of centrally
condensed, rotating, magnetised cores \citep{boss97}. More complicated hypotheses appeal to the
chaos of the cloud environment, such as core-core collisions
\citep{turneretalcap95}, protostellar encounters \citep{shenetal10}, dynamical 
capture in unstable multiple systems \citep{batebonnellbrommbin02}, and
accretion-triggered fragmentation (\citealt{bonnell94};
\citealt{whitworthetalrot95}; \citealt{hennebelleetalcomp04};
\citealt{offnerklein08}; \citealt{kratter_2010}).

Star formation scenarios involving the rapid collapse of a
protostellar core following the loss of support against gravity
require the formation of a massive accretion disc around the central
object (e.g. \citealt{vorobyovbasu07} and \citealt{walch09}). Such
discs undergo a short-lived $\sim 0.1\rm\,Myr$ stage where the disc is
massive relative to its host ($0.1 < M_d/M_{\star} < 1$) and where
gravitational instabilities operate to transport mass through the disc
onto the central protostar \citep{vorobyovbasu07}. Massive, accreting,
and extended protostellar discs have been shown to be susceptible to
fragmentation, which could be responsible for a range of phenomena
such as FU Orionis events and early dust processing if the clumps are
disrupted (\citealt{boleyetal10}; \citealt{nayakshin2010}), and the formation of substellar
companions otherwise (\citealt{vorobyovbasu07} and
\citealt{boleytwomodes09}).

Given the problem that both the standard core accretion planet
formation timescale and protostellar disc lifetimes are typically a
few Myr \citep{haischladalada01}, the idea of creating giant planets
in a few orbital times via gravitational instabilities was revived
\citep{bosssci97} and has since been the subject of sustained interest
(\citealt{boss02,boss08}; \citealt{mayeretal04,mayeretal07};
\citealt{pickettetal00}; \citealt{pickett07}; \citealt{boleyetal06};
\citealt{boleytwomodes09}) . Analytical works constraining inner disc
fragmentation \citep{rafikov05,rafikov07}, and the short cooling times
required to form long-lived clumps within $\sim 10\rm\,AU$ in
simulations of protoplanetary discs, along with observations of
massive planets on wide orbits (e.g. \citealt{maroisetal08,maroisetal2010}), have
lead to a shift in focus to outer disc ($> 40\rm\,AU$) fragmentation
(e.g. \citealt{stamatellosfrag07}; \citealt{boleytwomodes09}; \citealt{dodson2009}; 
\citealt{vorobyovbasu2010}). Whether the extended disc models
used in these studies are similar to discs formed from collapsed
molecular cloud cores remains to be seen.

In light of the likely connection between outer disc fragmentation and
early protostellar systems, we examine in this paper the formation of
discs in detail using 3D SPH simulations of core collapse in turbulent
molecular clouds. In particular we compare the early evolution in
mass, surface density, specific angular momentum, and disc stability
between several different systems under near-identical conditions. In 
one set of simulations we compare an
isolated and a binary system, and in the other we compare an isolated
system and a small cluster. Previous studies of gravitational
instabilities in binary systems have yielded mixed results, with some
finding that the perturbing companion hinders fragmentation through
disc truncation and tidal heating (\citealt{nelson00};
\citealt{mwqs05}) and alternatively promotes fragmentation, also
through tidal perturbations \citep{boss06}. However, the evolved
systems considered in these studies may not be as susceptible to
fragmentation as their protobinary counterparts, which we investigate here,
 due to the enhanced importance of gravitational instabilities during
protostellar disc formation.

The paper is laid out as follows: in Section 2 we discuss the
simulations, initial conditions and the clump identification procedure. In
Section 3 we present the results and analysis. Further discussion is provided in
Section 4, and our conclusions are laid out in Section 5.

\section{The Simulations}
All calculations were run with Gasoline \citep{gasoline04}, a parallel implementation of TreeSPH. 
We employ a fixed number $N$ of smoothing neighbors, with the main runs using $N = 32$. 
Artificial viscosity is the standard prescription \citep{monaghanvisc83}
with $\alpha = 1$, $\beta = 2$, controlled with a Balsara switch \citep{balsaraphd89}. 

\subsection{Initial Conditions}

The cloud in this experiment is spherical and uniform with a radius of~$\rclf$,
a mass of~$\mclf$, and a temperature of~$\tcl$, as in~\citealt{bateclbd03}. It was
seeded with supersonic turbulent velocities and is marginally self-bound. To
simulate interstellar turbulence, the velocity field of the cloud was generated
on a grid as a divergence-free Gaussian random field with an imposed power
spectrum $P(k) \propto k^{-4}$. The resulting velocity dispersion $\sigma(l)$
varies as $l^{1/2}$, and is consistent with the Larson scaling
relations~\citep{larson81}. The velocities were then interpolated from the grid
to the particles.  Finally, the condition that the cloud be marginally
self-bound gives a normalization for the global velocity dispersion
of~$\sigclf$. We describe the results of two simulations in this paper, \rname{A} and
\rname{B}. 
\Rnames{A \& B} are different random
realizations of the same ICs, with the same power spectrum, but with
different initial phases of the velocity field. As a result the first objects
that form in \rname{A} (which are the only ones that we can follow
owing to the high computational expense of modeling hydrodynamics at
high resolution without employing sinks), are different from those in
\rname{B}.
In \rname{A} we have two sub-simulations, one in which a binary system naturally arises, and another
in which we suppress the production of the secondary. We then compare the subsequent 
evolution of the systems in an otherwise identical core and cloud environment. 
In \rname{B} an isolated system and a cluster are produced roughly coevally at opposing ends of a 
filamentary structure. We compare the evolution in these systems, having been produced
in similar environments.

\subsubsection{\Rname{A}: The Isolated and Binary Systems}
We describe here how we produced the isolated system in \rname{A}.
The fiducial cloud free fall time is given by~$\tffs = \tffm = \tfff$, with 
$\rho_{cl} = \rhocl$ being the initial cloud density. 
The evolution of the cloud was followed up to~$t = \tsim$.

At $t = \tstir$ a core, whose collapse we follow in detail, has already formed a 
small central protostar. The protostar accretes rapidly via a connecting 
filament until the filament itself produces a fragment. The fragment survives
first pericenter with the protostar and becomes a companion.

We produced an isolated system for comparison under near-identical conditions
by tracing the particles comprising the companion back in time to~$t = \tstir$ 
and stirring them by randomly exchanging their velocities. This
procedure conserves the energy and linear momentum but not the
angular momentum of the stirred particles. There were approximately $5300$ particles,
totaling $\msl{0.053}$, that had their velocities perturbed. 

\subsection{Thermodynamics}

We model the thermodynamics of collapse with a piecewise polytropic equation of state 
(\citealt{tohline82} and~\citealt{batestar98}):
\begin{equation} \label{eqn:pp}
  \pqtrop
\end{equation} 
with $\kappa_0$ chosen so that the sound speed $c_s = \csnaught$ and the subsequent 
$\kappa_1$ chosen to ensure pressure continuity. Theoretical estimates of
temperatures in molecular clouds find that their 
temperatures should range between 5--10\,K over densities ranging from
$10^{-19}$--$10^{-13}\gcc$ (\citealt{larson85};
\citealt{lowlyndenbell76}; \citealt{masunagainutsuka00}). Observations
indicate a somewhat higher minimum temperature of 8\,K, with typical
temperatures of 10--13\,K \citep{kirkwardthompsonandre07}. The equation of
state thus captures the approximately isothermal behaviour of the cloud in the
intermediate density regime. At high densities the gas transitions to being
adiabatic with exponent $\gamma = 7/5$. This equation of state
is a simplification of the internal heating and cooling processes in a
molecular cloud and, in the high density regime, will minimize the
potential for fragmentation. Furthermore, through our choice of softening ($\epsilon_g = \soft$) combined with
the polytropic equation of state, we allow the first hydrostatic core to be
marginally resolved,  while still inhibiting its subsequent dissociating
collapse \citep{larson69}, which can be computationally demanding to resolve.
Three-dimensionsal collapse simulations
with more sophisticated thermodynamics, and including radiative transfer, but of
isolated cores with no turbulent cloud environment, have been performed by 
(\citealt{whitehousebate06}; \citealt{batestar10};
\citealt{tomidaetal10}; \citealt{commerconetal10}). While proper
treatments of both thermodynamics and radiative transfer are important,
such simulations are computationally expensive and so far can only be carried out
at lower resolution, and with shorter integration times than required in our
calculations.  We shall explore the problem with more detailed thermodynamics
and radiative transfer in a future publication. We reiterate, though, that the early
evolution of protostellar discs and protobinaries is still not
understood.  Before the complexities of radiation hydrodynamics can be
introduced, the birth of these systems must be understood through
examination of high-resolution hydrodynamics simulations with a reasonable,
albeit simplified, EOS, which is the focus of this paper.

\subsection{Resolution}
The simulations employ $5\rm\,M$ particles. The particle masses $m_p$
are thus $\msl{1.0\times 10^{-5}}$ in all simulations. The minimum jeans mass $M_j$, defined as $M_j =
2.92\,c_s^3/(G^{3/2}\rho_0^{1/2})$, is that obtained at the transition
density to the adiabatic equation of state, and is $M_j^{*} =
\msl{1.7\times 10^{-3}} = 1.8\rm\,M_{jup}$. Previous work has shown
that molecular clouds should have a minimum jeans mass
\citep{lowlyndenbell76}, with the gas becoming optically thick to its
own cooling radiation at $\sim 10^{-13}\gcc$.  In our $5\rm\,M$
particle runs $M_j^{*}/m_p = 167$ and the jeans mass remains
adequately resolved at all times. A fixed gravitational softening
$\epsilon_g = \soft$ is used in all simulations.

\subsection{Disc Identification}
\label{sec:discid}
In order to compare and separate the systems in the 
simulations, a working definition of what comprises a protostellar system is
required. We define a protostellar system within this context as being a
self-bound gas structure with a peak density greater than $\rhothresh$. 
The density threshold was chosen to be well above the critical density
in order to identify only bound structures that are in the adiabatic
regime. To determine whether any cold, isothermal gas is bound to a given system, we require that 
it is bound below a threshold binding energy $E_b = \ebind$. The binding 
energy criterion is imposed in
order to exclude particles that are only marginally or not uniquely bound
to the system, e.g., lying in the protostellar envelope or flowing between the primary and
secondary. The value was chosen empirically to be as small as possible while
minimising noise in the results. 

To put this into practice we use the SKID \citep{stadelphd01} group finder. 
SKID works by pushing tracer particles along density gradients to find local 
maxima, and linking them with friends-of-friends. Once group assignments are found we then
remove unbound particles to create self-bound groups, computing:
\begin{equation} \label{eqn:eskid}
  E = \frac{m_p}{2}|{\bf v}-{\bf v}_{cm}|^2 + U + E_{th}
\end{equation} 
as the total energy, with ${\bf v}$ and ${\bf v}_{cm}$ the velocity and center of mass velocity,
$U$ the gravitational potential, and $E_{th}$ the thermal energy. The center of mass frame is updated 
throughout the unbinding procedure. 

\section{\Rname{A}: Results}

\begin{figure}
\includegraphics[width=\fw\textwidth]{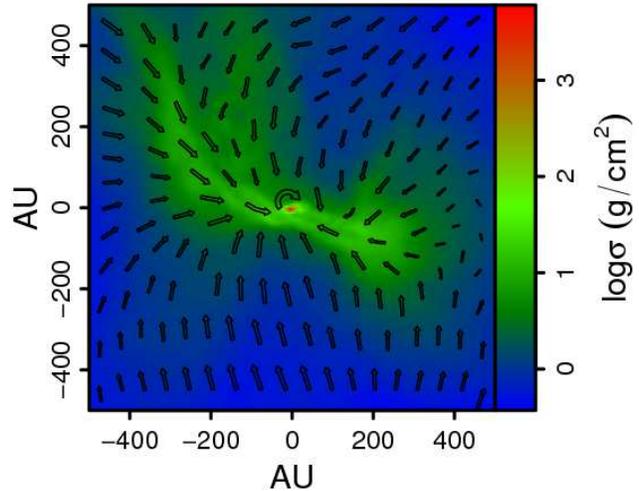}
\caption{ The column density and velocity field in a $(1000{\rm\,AU})^3$ volume,
containing~$\msl{0.29}$ of gas, around the central prestellar object. The time
is $t = \tstir$, at the onset of core collapse. }
\label{fig:lscale}
\end{figure}

\begin{figure}
\includegraphics[width=\fw\textwidth]{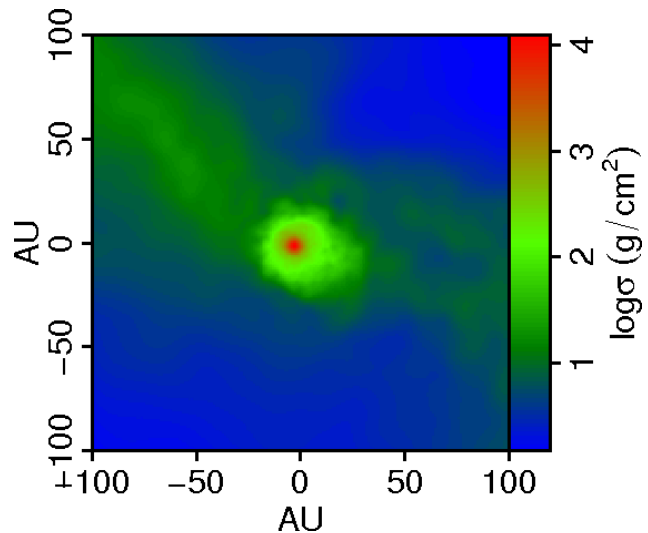}
\caption{ The protostellar object formed at the onset of its parental
  core's collapse at $t = \tstir$. 
  The object is $\tfi$ old at this stage with a disc of $25{\rm\,AU}$ in radius, and a mass of $\msl{0.055}$. }
\label{fig:inset}
\end{figure}

At $t = \tstir$ a nascent protostellar system has begun to form from
the collapse of its enveloping core. In Fig.~\ref{fig:lscale}, the
surface density and velocity field, projected through a
$(1000{\rm\,AU})^3$ box cut around the prestellar object, are shown.
The box contains~$\msl{0.29}$ of gas. The central prestellar object
formed via fragmentation of a filament $1000{\rm\,AU}$ in extent.  We
see in the integrated transverse velocity field that accretion
primarily occurs along the filament, as there is only a small torus of
opening angles around the system from which low density material is
observed to make a direct approach. Most gas collides with the
filament and is subsequently funneled onto the system.

In Fig.~\ref{fig:inset}, the surface density of the prestellar system,
in projection along its axis of rotation, is shown.  The system is
$\msl{0.055}$ and the disc is $25{\rm\,AU}$. The filaments feeding the
disc at this stage have densities ranging from $\rhofill$ up to the
critical adiabatic density $\rhofilh$. The system accretes rapidly
until $t = \tbinff$ whereupon a fragment of mass $\mfrag$ forms in the
low density filamentary material, close to the Jeans mass of
$\mjfrag$.  The fragment survives first pericenter and becomes a
companion. The isolated system, in contrast, continued accreting
material rapidly, building up the disc around the protostellar system.

\subsection{Surface densities and temperatures}

\begin{figure*}
\centering
\includegraphics[width=\fw\textwidth]{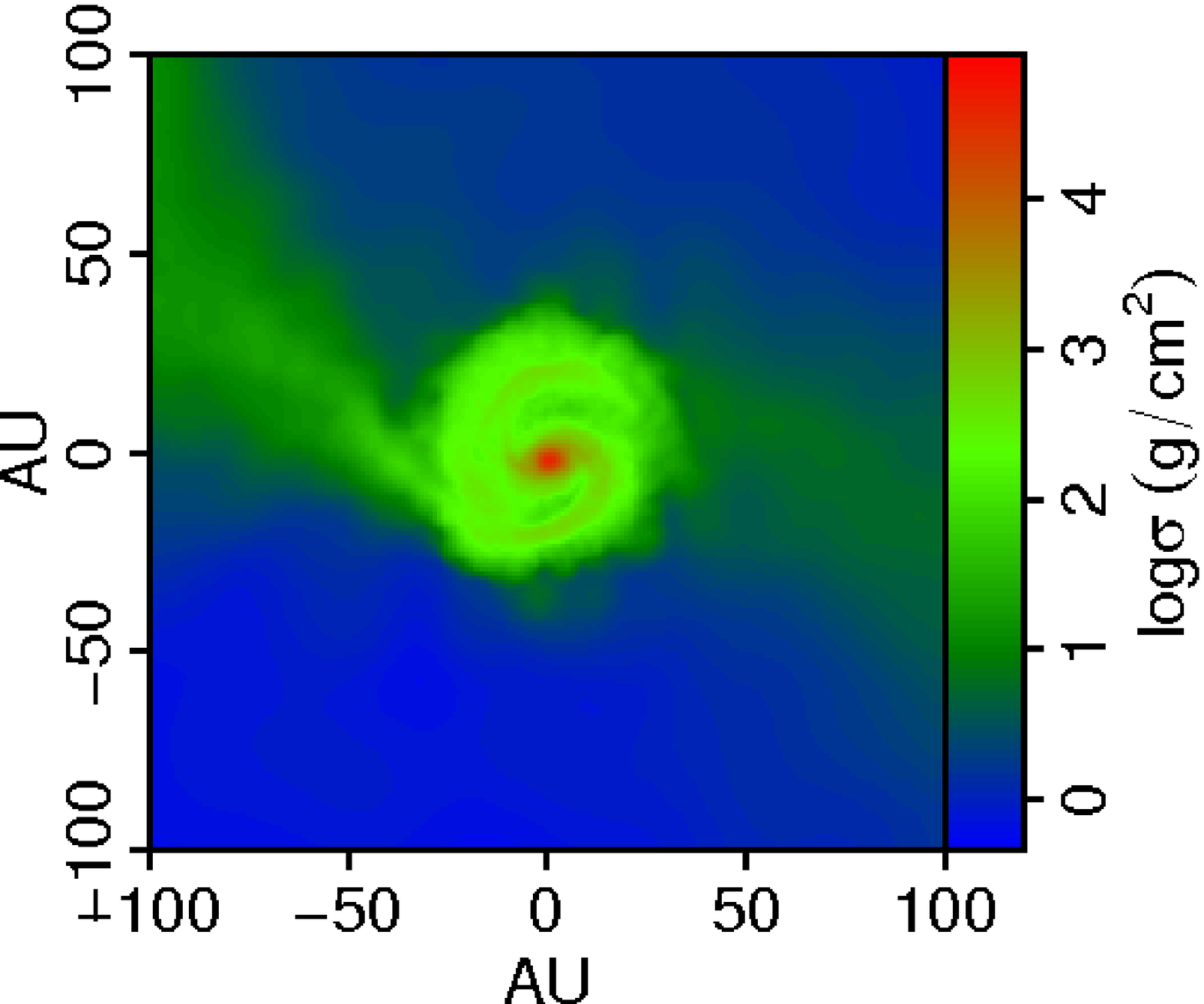} 
\includegraphics[width=\fw\textwidth]{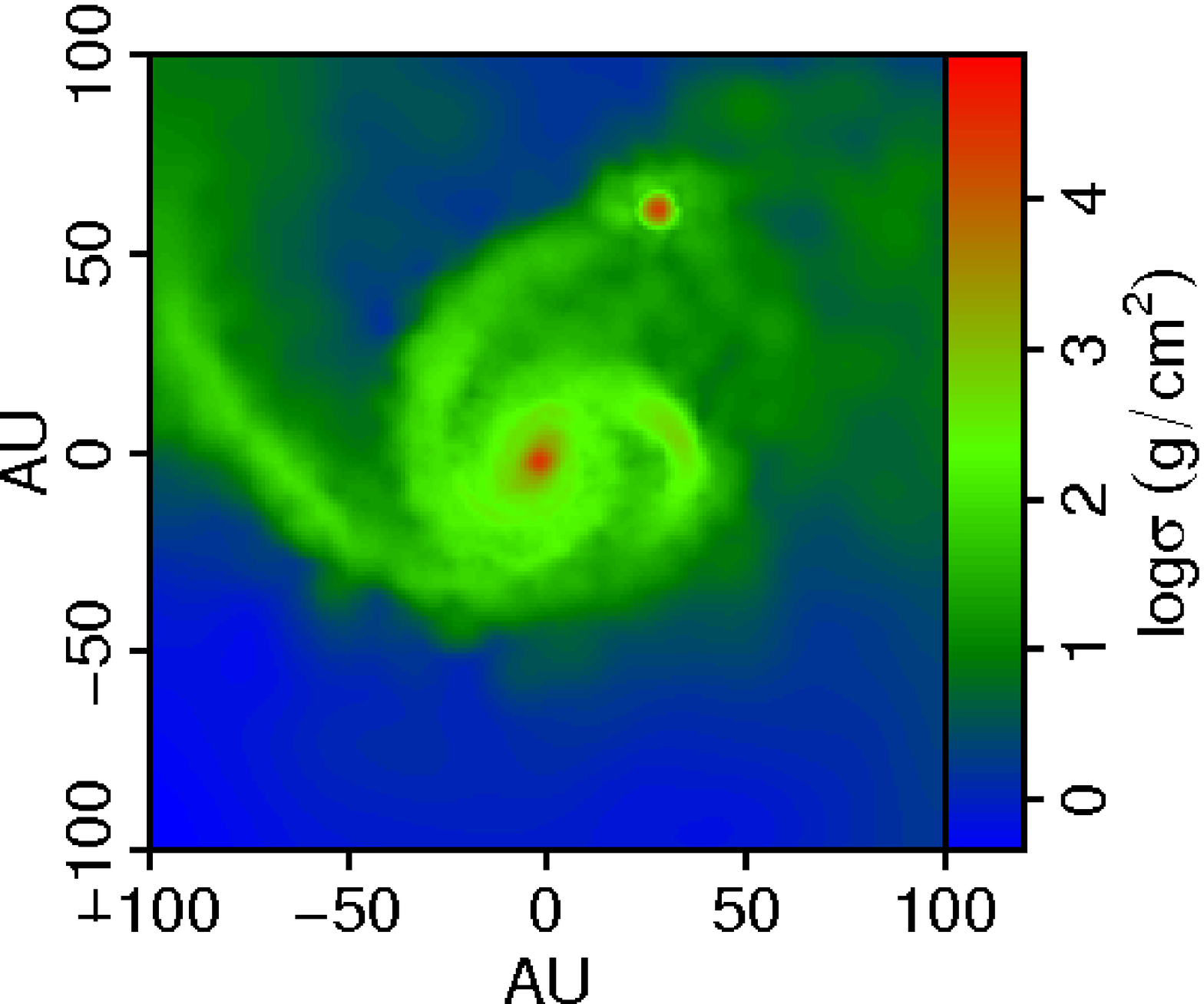}

\includegraphics[width=\fw\textwidth]{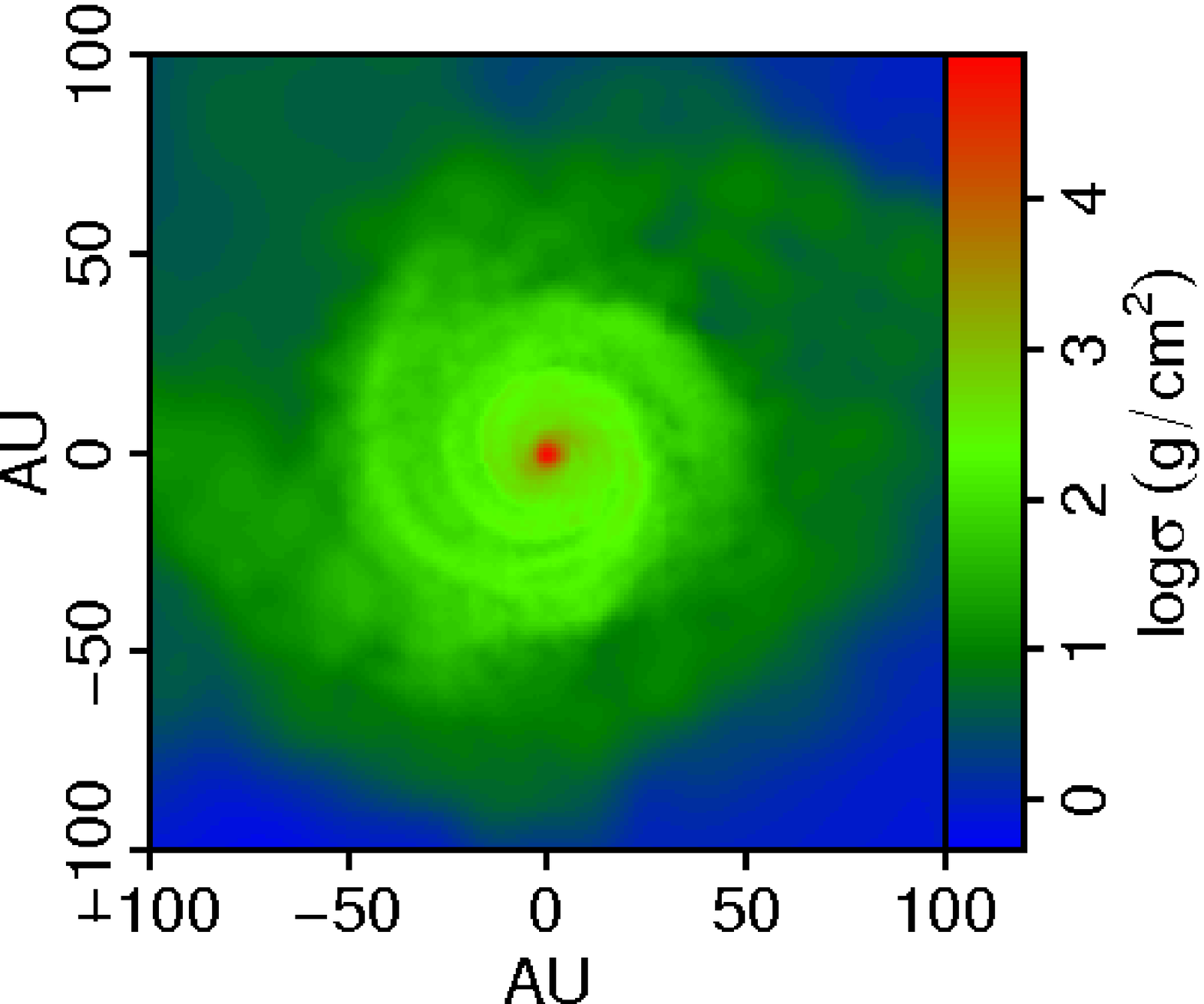}
\includegraphics[width=\fw\textwidth]{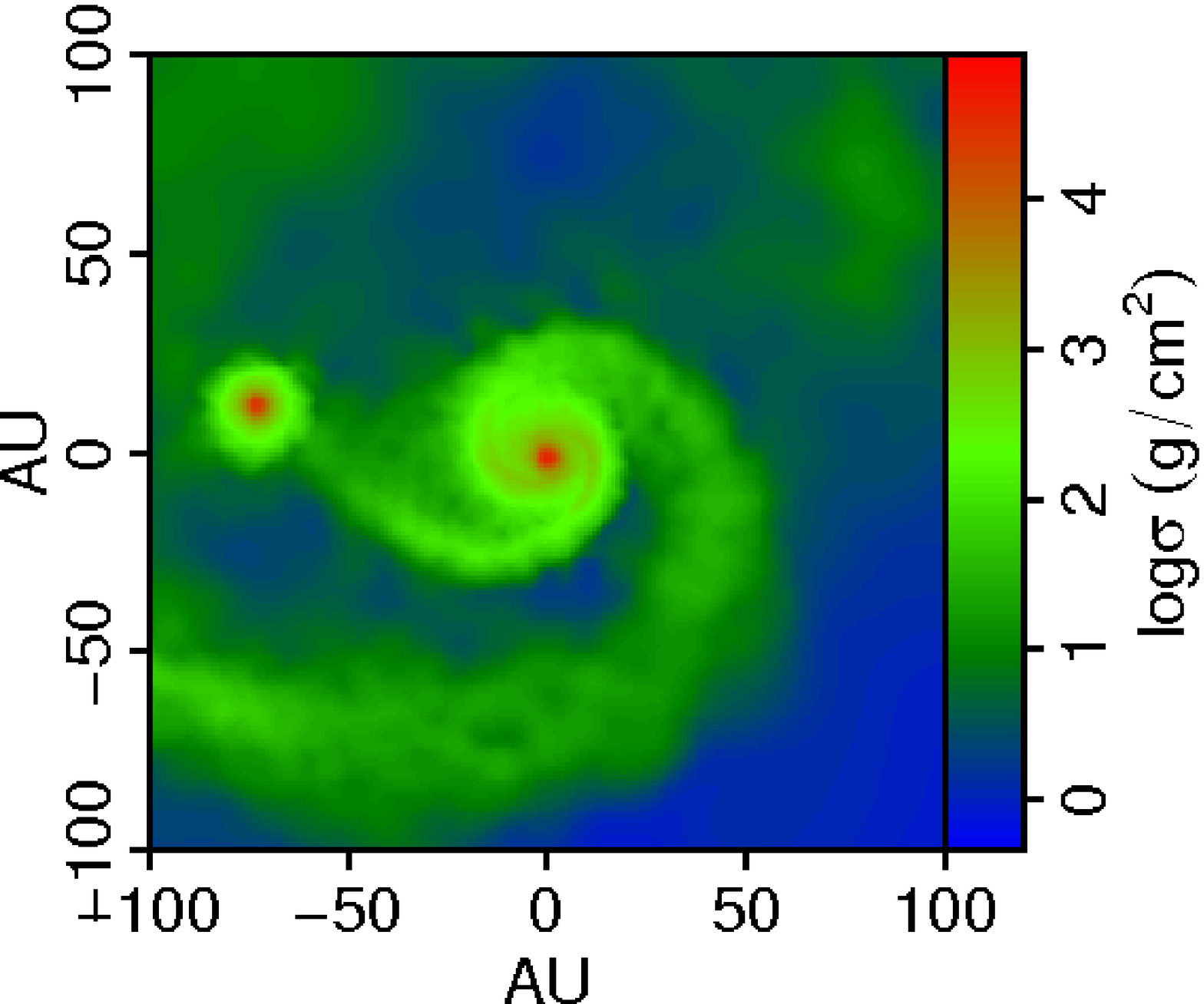} 

\includegraphics[width=\fw\textwidth]{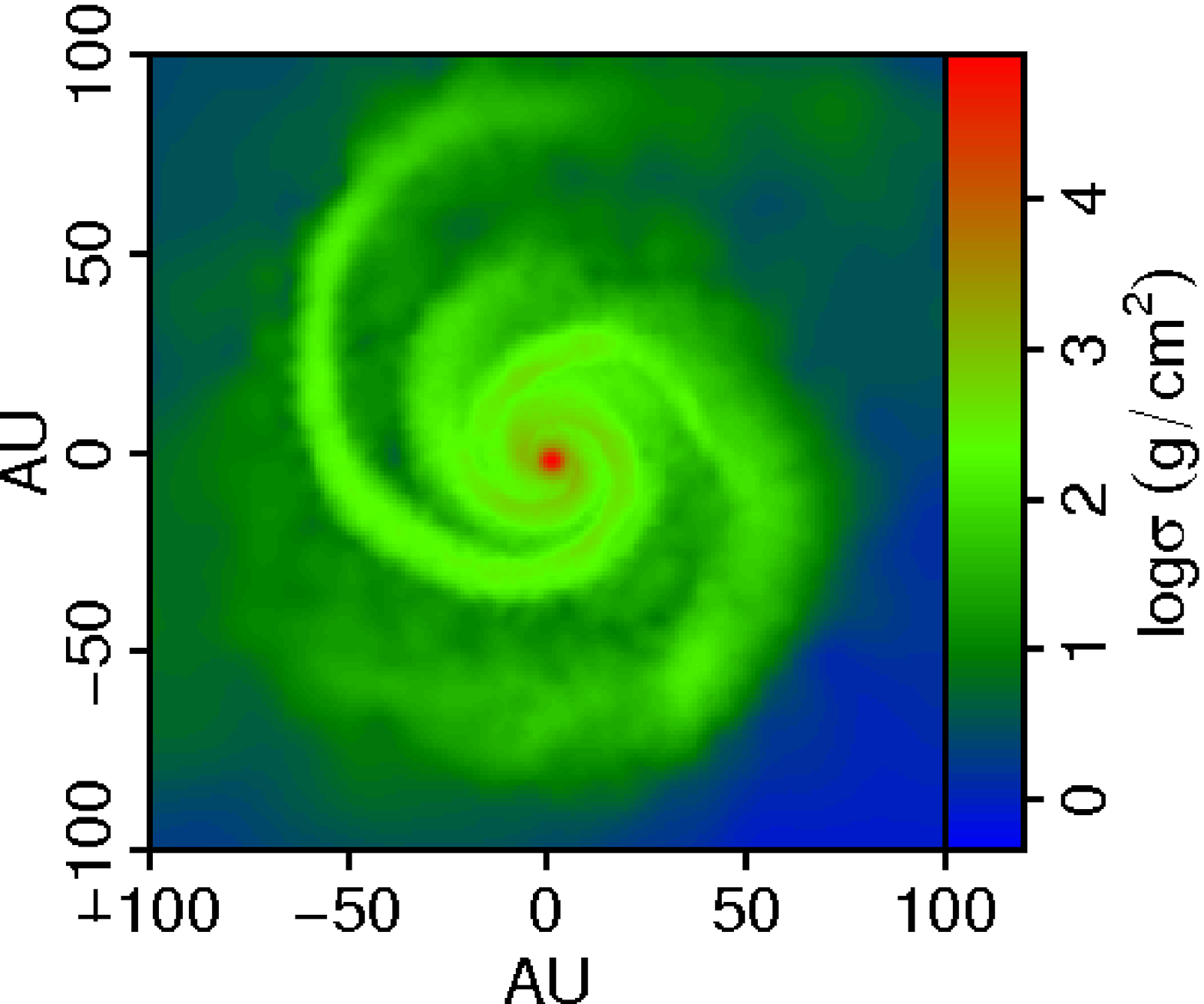}
\includegraphics[width=\fw\textwidth]{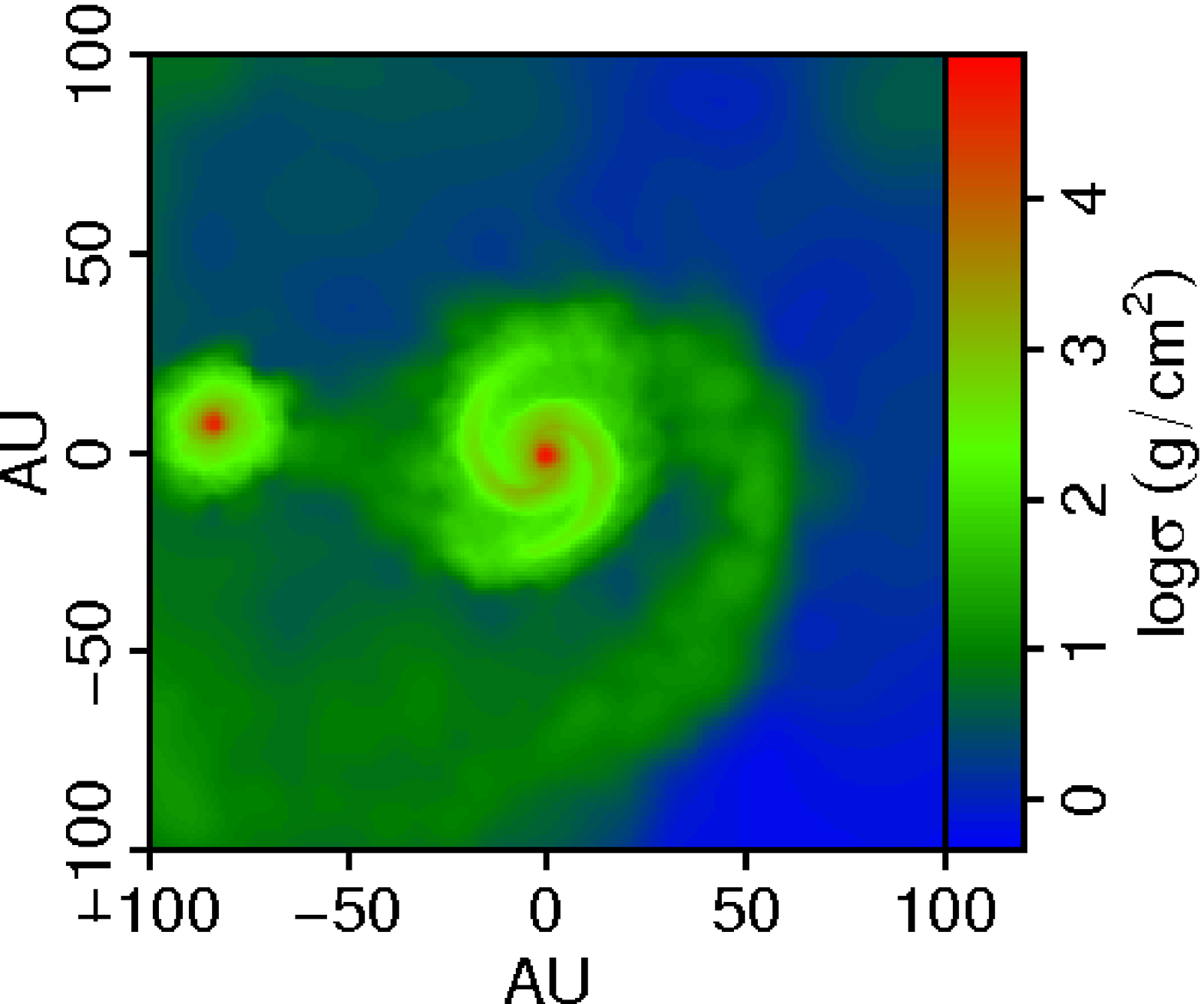} 
\caption{ The surface densities of the isolated and binary systems, respectively in the left and right columns,
    and shown at times $\tpa$, $\tpb$, and~$\tpc$ from the top frow to bottom, after formation.}
\label{fig:sigpanel}
\end{figure*}

\begin{figure*}
\centering
\includegraphics[width=\fw\textwidth]{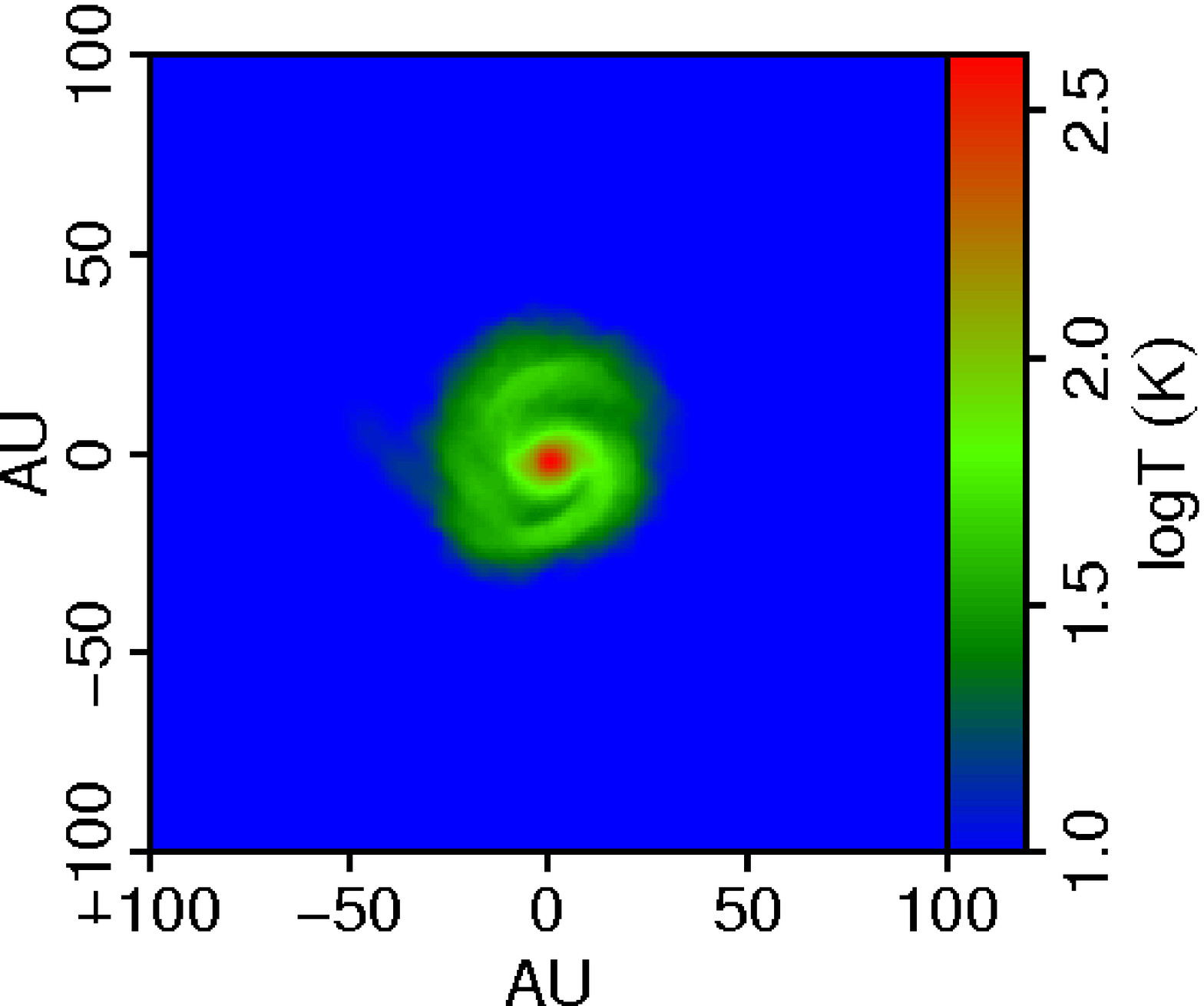} 
\includegraphics[width=\fw\textwidth]{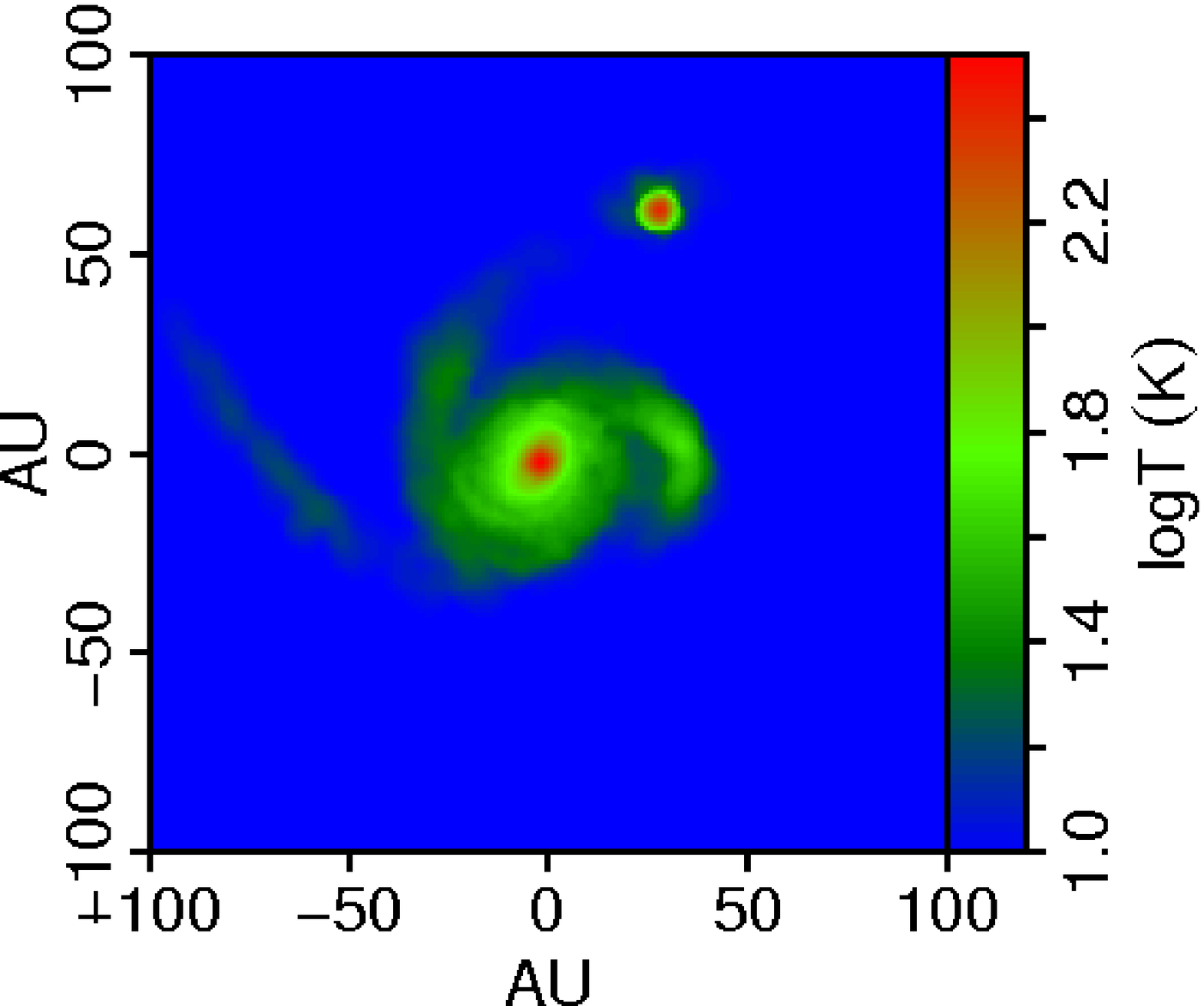}

\includegraphics[width=\fw\textwidth]{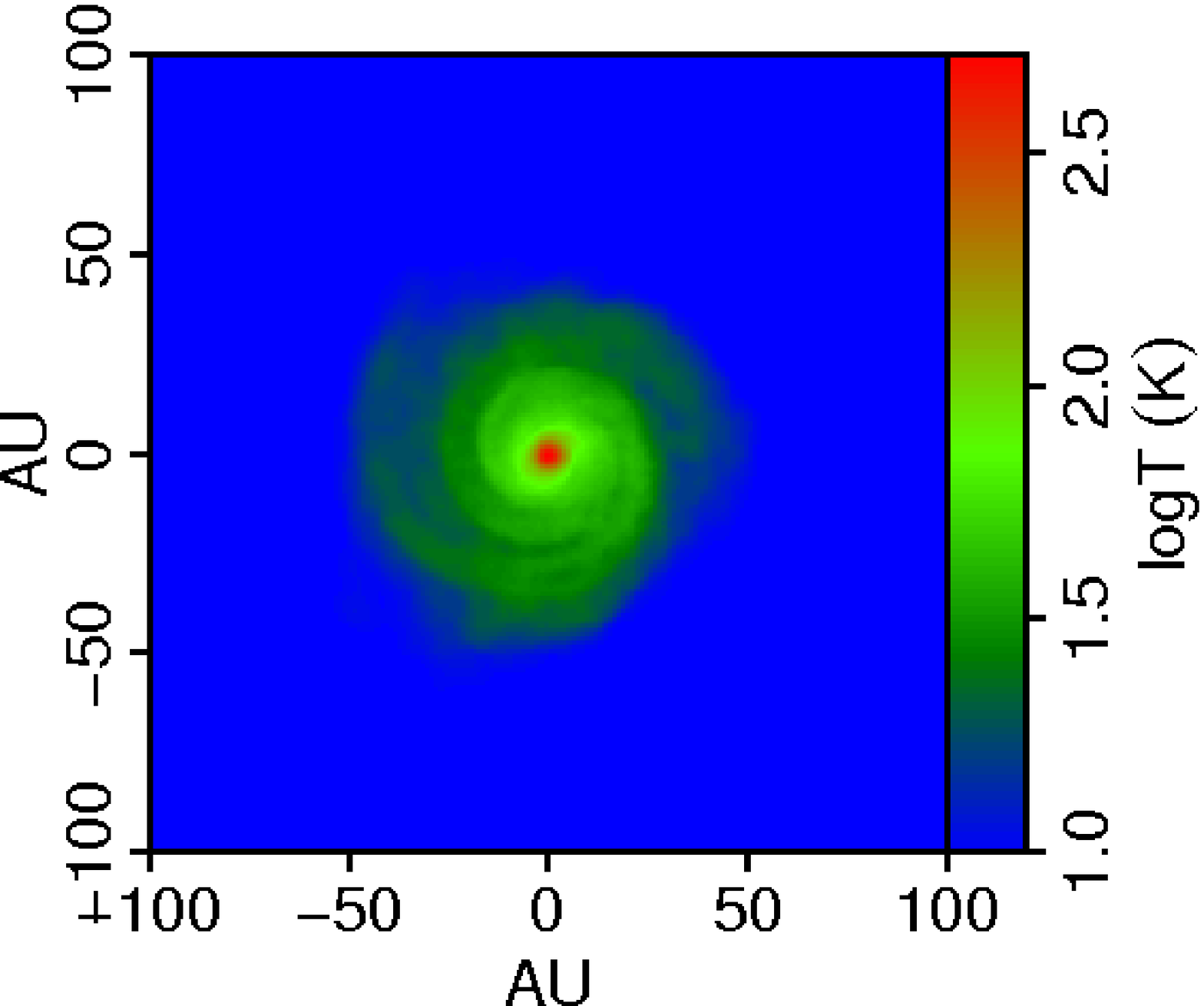}
\includegraphics[width=\fw\textwidth]{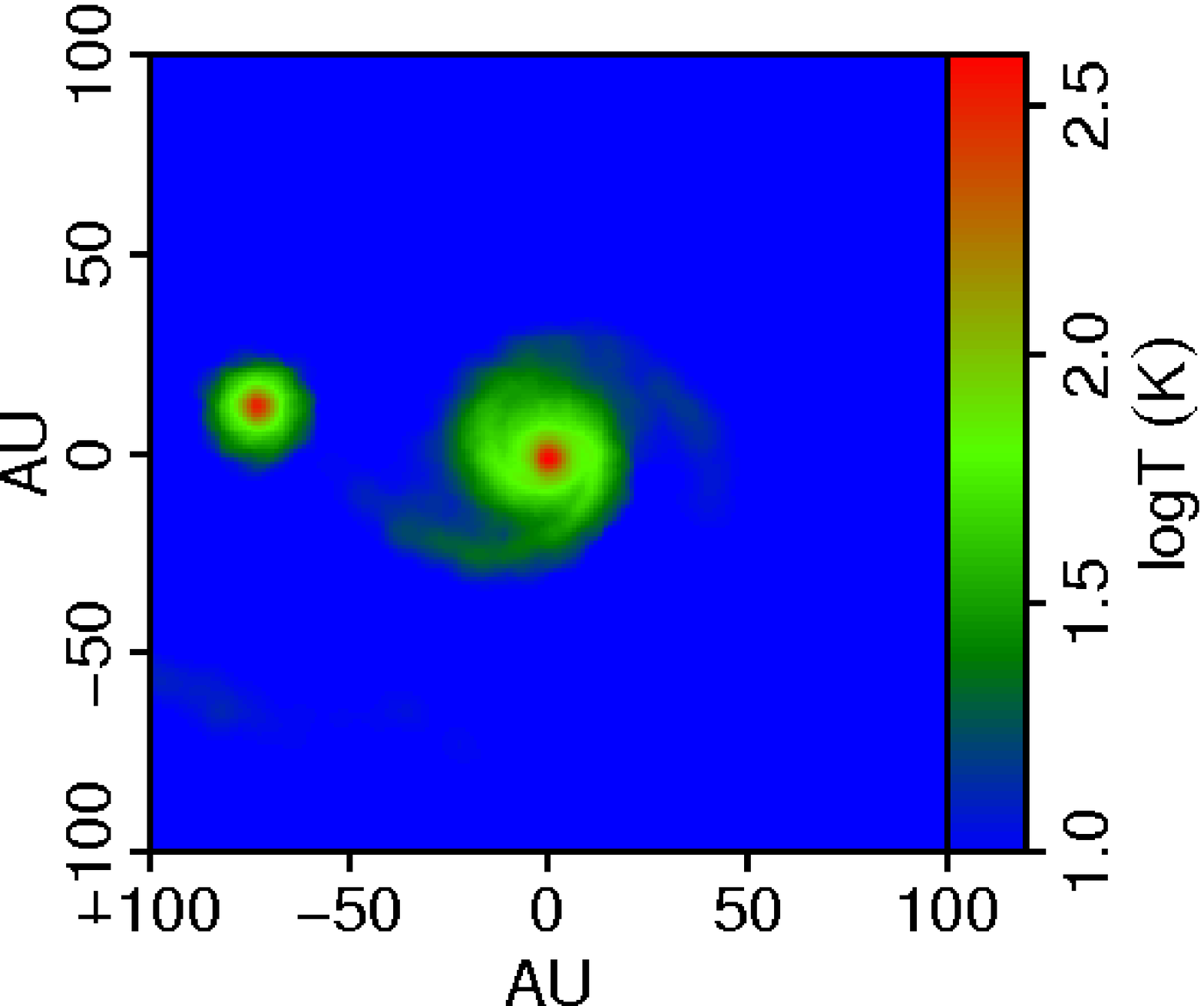} 

\includegraphics[width=\fw\textwidth]{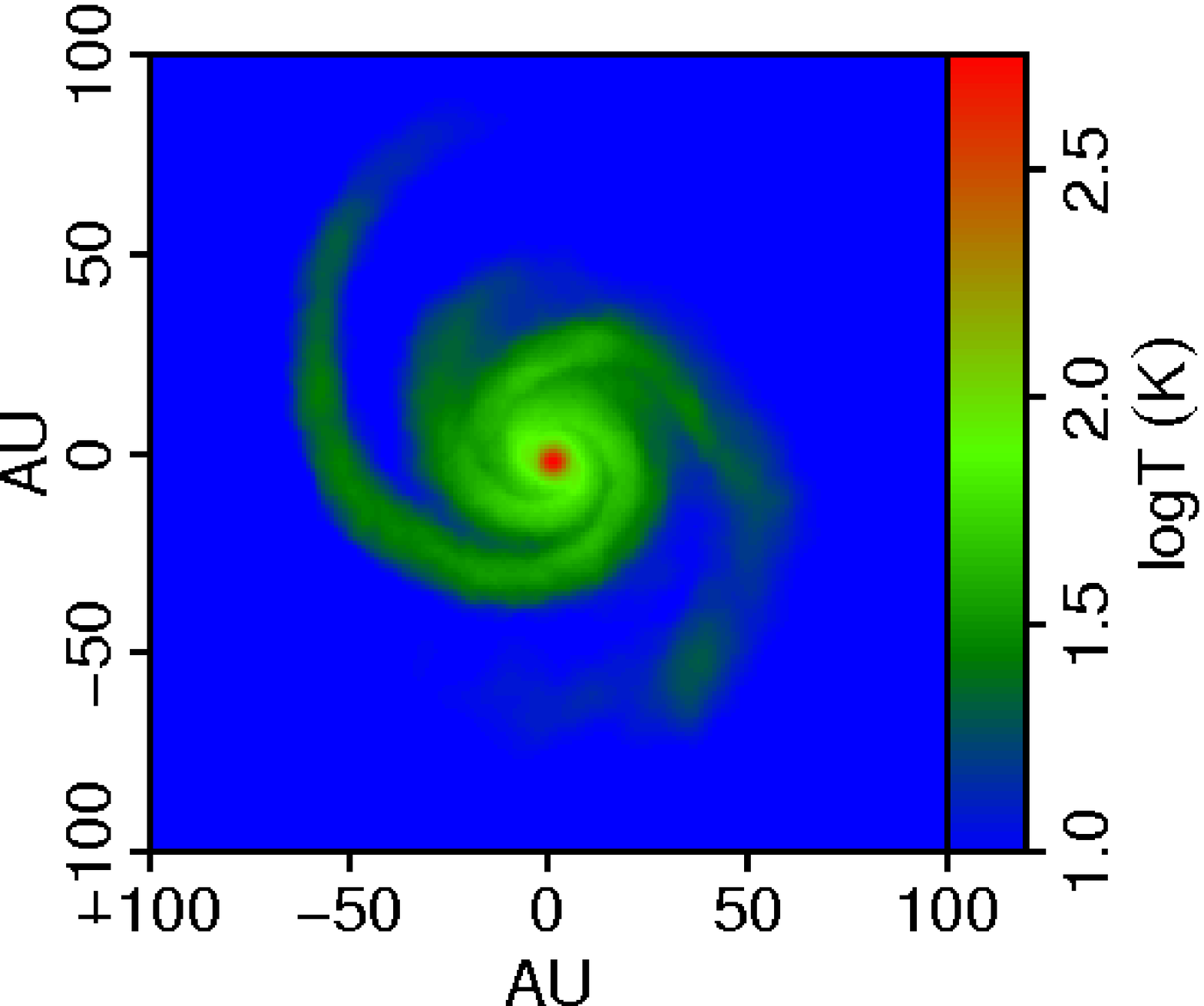}
\includegraphics[width=\fw\textwidth]{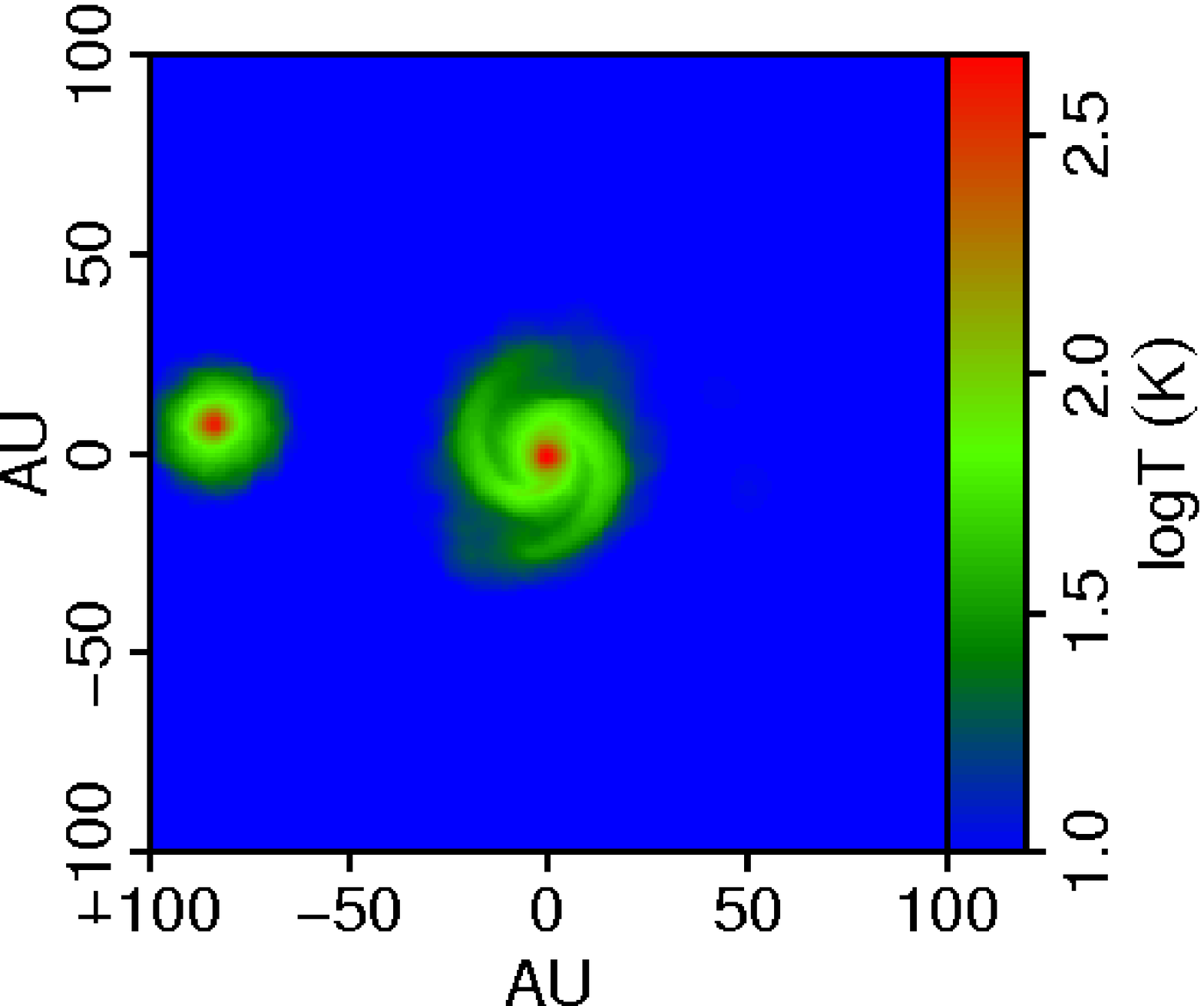} 
\caption{ Mass-weighted temperature maps of the isolated and binary systems, respectively in the left and right columns,
    and shown at times $\tpa$, $\tpb$, and~$\tpc$ from the top frow to bottom, after formation.}
\label{fig:tempanel}
\end{figure*}

\subsubsection{Surface Density Maps}
In this section, we compare the divergent evolution of the isolated
and binary systems. To accommodate the change in dynamical times at
disc scales we set the units to kyr and count time relative to the
formation of the system at~$\tbirth$. In Fig.~\ref{fig:sigpanel} we
see the surface density maps of the isolated and the binary systems in
the left and right columns respectively, shown at times $\tpa$,
$\tpb$, and~$\tpc$ after formation of the primary (or equivalently,
the isolated system). Differences appear early between the two
systems. In the topmost row, we see that the primary in the binary
system is similar to the isolated system in morphology and surface
density except for a tidal disturbance of the disc caused by an
initial close passage of the secondary. In the central row the
differences are more pronounced. At $\tpb$ we see that the isolated
system has continued to grow in mass and the disc now has a radius of
$50{\rm\,AU}$. Between $\tpa$ and $\tpb$ approximately 1 orbital time
has passed in the binary system and the secondary is at its
apogee. The primary has shrunk in extent as the secondary has accreted
some the gas in its outer regions.  The secondary is also rapidly
accreting high specific angular momentum gas from the two filaments
feeding the system.  In the bottom row at $\tpc$ we see that the
isolated system has continued to grow in mass and extent but has
become gravitationally unstable and an $m = 2$ spiral arm has
developed into a material arm. At $\tpc$ the binary system has
completed another orbit, both systems have grown in mass, and the same
pattern of tidal disturbance of the primary is observed.

\subsubsection{Temperature Maps}
In Fig.~\ref{fig:tempanel}, mass-weighted temperature maps of the
binary and isolated systems, at times $\tpa$, $\tpb$ and $\tpc$, are
plotted, as in Fig.~\ref{fig:sigpanel}. As we are using a piecewise
polytropic equation of state (Equation~\ref{eqn:pp}), some of the gas
bound at the outer edge of the system(s) and some interarm gas is
still in the isothermal phase. In the isolated system, increasing
amounts of mass end up at large radii and at low densities, due to its
higher specific angular momentum $j$ (see section~\ref{devo} for elaboration
on these points).  In the binary system however, pictured in the right
column of Fig.~\ref{fig:tempanel}, almost all of the mass identified
as bound to each system lies at high densities (and is thus adiabatic)
and within their respective tidal radii.

\subsection{Disc Evolution} \label{devo}

\subsubsection{Disc Mass}

\begin{figure}
\includegraphics[width=\fw\textwidth]{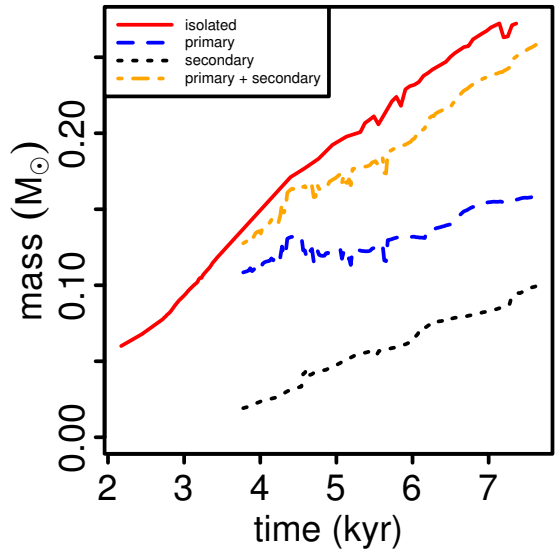}
\caption{ Total mass accreted (disc + protostar) versus time for the isolated, 
primary, and secondary systems, and the net mass in the binary system. }
\label{fig:mdisct}
\end{figure}

\begin{figure}
\includegraphics[width=\fw\textwidth]{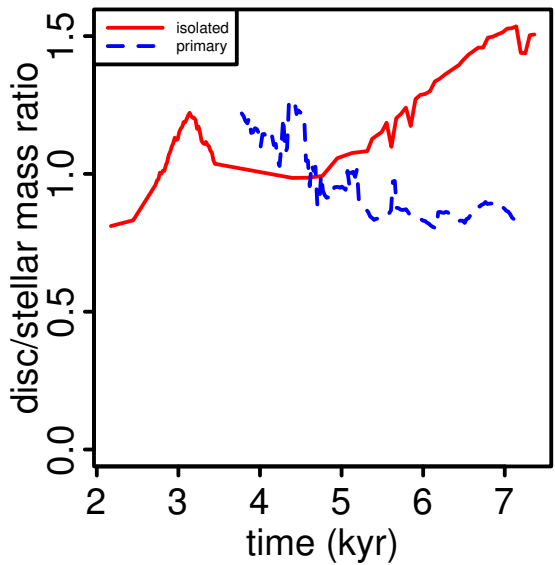}
\caption{ The ratio of disc mass to accreted stellar mass versus time, for the
  isolated, primary systems. }
\label{fig:mrat}
\end{figure}

In Fig.~\ref{fig:mdisct}, the accretion histories of the isolated system,
primary, and secondary are plotted. The mass at a given time is determined by
the procedure outlined in ~\ref{sec:discid}, and is thus the combined
protostellar and disc mass.  The isolated system accretes gas steadily and at a
high rate throughout the simulation, with typical accretion rates of $\mslyr{3-5
\times 10^{-5}}$. After the initial collapse, the central protostar, defined as
the mass contained within $r < 5\rm\,AU \simeq 2\epsilon$, has an accretion rate
of $\mslyr{5-6 \times 10^{-6}}$. Taking the asymptotic rate
from 1D protostellar collapse theory, $\dot{M} = m_\circ c_s^3/G$
\citep{stahlerpalla05}, we find that $\dot{M} = \mslyr{2 \times 10^{-6}}$, and so
 $m_\circ$, a constant of order unity, is approximately $2.5$ at the protostar. This is
not a high accretion rate, as 1D collapse simulations asymptote to $m_\circ = 2$
\citep{stahlerpalla05}. The accretion rate onto the entire system however, is
high, with $m_\circ \sim 10$, and is perhaps due to embedding within a
filament. Both accretion rates are consistent with other studies, such as those
found in \citealt{walch09}. 

As expected, the net mass accreted by the binary system is close to that of the isolated
one. The difference between the two is because there is some mass present in the
binary that can not be uniquely assigned either to the primary or
secondary. The binary system is plotted starting at $3.8\rm\,kyr$, instead of
its formation time at~$\tbinan$, because during the first pericenter the
identification procedure has difficulties separating the two objects
uniquely. In the binary system the secondary, due to its orbit, preferentially
accretes high specific angular momentum gas from the filaments and accretes mass
at a higher rate than the primary, and at later times the mass ratio tends
toward unity. This behaviour has been noted before in the literature
(e.g. \citealt{batebonnell97}).

In Fig.~\ref{fig:mrat}, the ratio of stellar mass (the mass within
$5\rm\,AU$, chosen to correspond with the limits of our softening)
to disc mass for the isolated and primary systems are plotted. The isolated
system continues gaining disc mass faster than it can be
transported to the star, while the primary evolves steadily towards
stability. As systems normally become moderately self-gravitating at a mass
ratio of $\sim 0.1$, we see that these newly formed protostellar systems are in
a regime dominated by gravitational instability.

\subsubsection{Cumulative Mass Fraction and Mass Profiles}
\label{sec:cumass}

\begin{figure}
\includegraphics[width=\fw\textwidth]{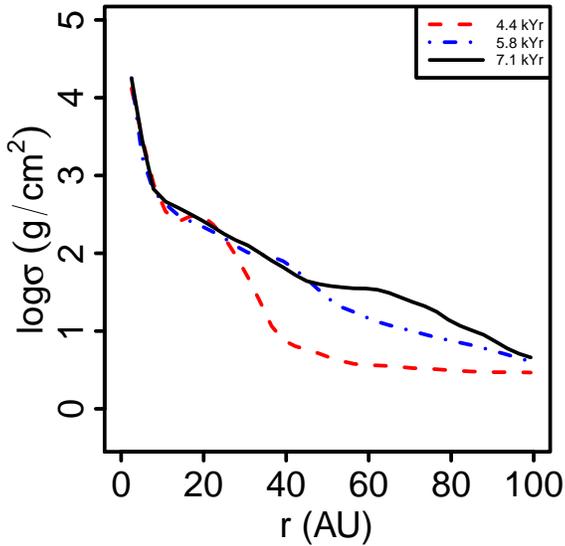}
\caption{ Evolution of the surface density profile of the isolated system. $\sigma$
          tends to increase at all radii with time. }
\label{fig:iso_sig1d}
\end{figure}

\begin{figure}
\includegraphics[width=\fw\textwidth]{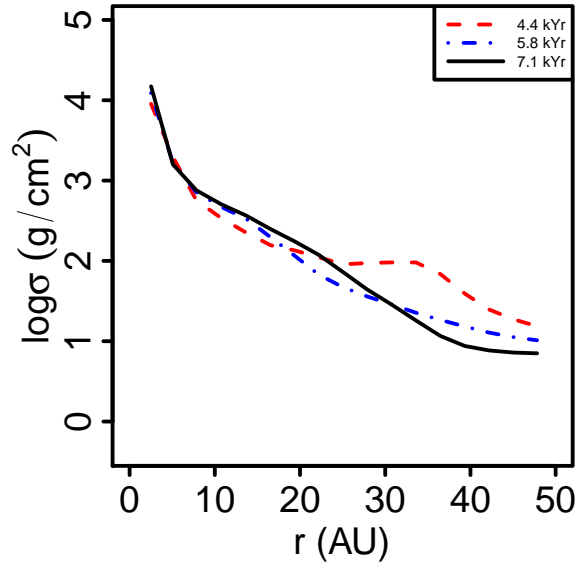}
\caption{ Evolution of the surface density profile of the primary in the binary system. 
          $\sigma$ decreases steadily at outer radii and increases at inner radii due to
          tidal interactions. }
\label{fig:bin_sig1d}
\end{figure}

\begin{figure}
\includegraphics[width=\fw\textwidth]{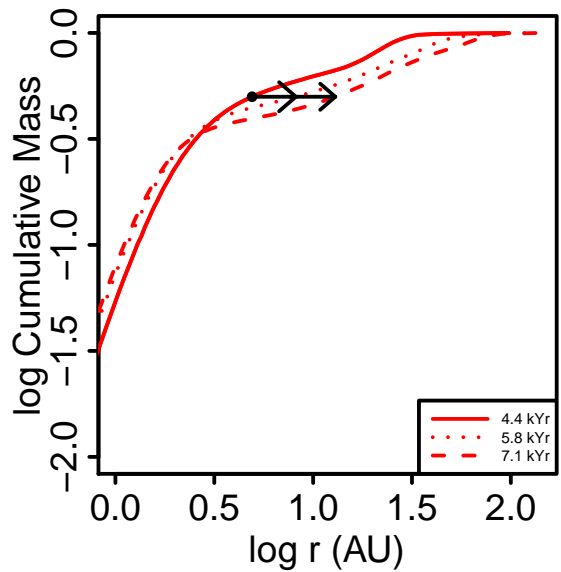}
\caption{ The mass profile evolution of the isolated system. As time progresses, 
          increasing amounts of mass relative to the total lies at large
          radii. We illustrate this by plotting the evolution of the half-mass radius in black.}
\label{fig:miso}
\end{figure}

\begin{figure}
\includegraphics[width=\fw\textwidth]{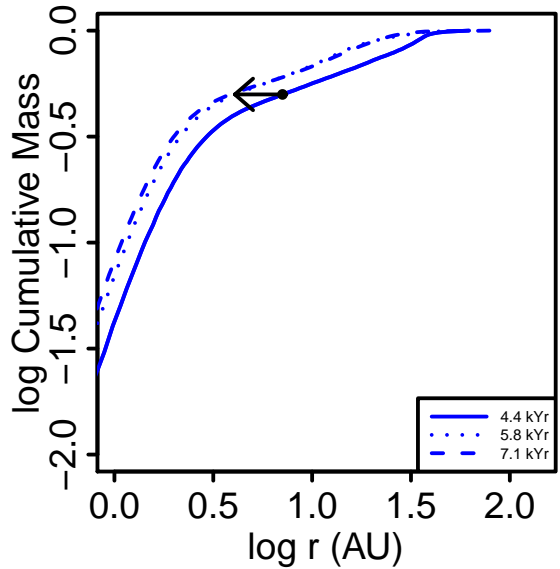}
\caption{ The mass profile evolution of primary. In contrast to
          Fig.~\ref{fig:miso} we see the relative fraction of mass at large radii
          decreasing with time. We illustrate this by plotting the evolution of the half-mass radius in black. }
\label{fig:mpprof}
\end{figure}

The discs in both the isolated and binary systems are all massive relative to
the central protostar and exhibit strong spiral arms as seen in
Figs.~\ref{fig:sigpanel}. We expect that tidal torques will, in addition, play a
role in triggering mass transport within the discs in the binary as noted for
example in \citep{mwqs05}. In Figs.~\ref{fig:iso_sig1d} \& \ref{fig:bin_sig1d}
are plotted the surface densities averaged in annuli of the isolated disc and of
the primary of the binary system. We see from Fig.~\ref{fig:iso_sig1d} that in
the isolated system the trend as time progresses is toward increasing amounts of
mass at all radii, while from Fig.~\ref{fig:miso} it is evident that the
relative amount of mass at large radii is increasing, illustrated by the outward
movement of the half-mass radius.  The surface density profile of the primary in
the binary system (see Fig.~\ref{fig:bin_sig1d}) shows increasing surface
density within $20\rm\,AU$ and clearly decreasing surface density outside of
$30\rm\,AU$ throughout the simulation. This is reflected in the cumulative mass
profile in fig.~\ref{fig:mpprof} as a decreasing overall fraction of mass found
at larger radii, and the clear inward movement of the half-mass radius (while
the total mass slowly increases). The source of this difference is likely
increased mass transport from gravitationally amplified tidal perturbations. To
illustrate the role of tidal effects we model the primary and secondary as point
masses and compute the Jacobi radius $r_j$ at $\tpb$ given the nominal values $M
= \msl{0.125}$, $m = \msl{0.05}$ and $R_0 = 75\rm\,AU$ (see
Figs.~\ref{fig:mdisct} \& \ref{fig:sigpanel}) for the primary mass, secondary
mass and separation, respectively. Given these parameters we compute $r_j = 0.41
R_0 = 31\rm\,AU$, thus we expect tidal effects to, at the very least, limit the
disc of the primary to $44\rm\,AU$. Looking at the cumulative mass profile, we
see that at our disc-finding procedure is roughly consistent with such a simple
model, with very little bound mass found outside of $30\rm\,AU$.

\subsubsection{Temperature Profiles}

\begin{figure}
\includegraphics[width=\fw\textwidth]{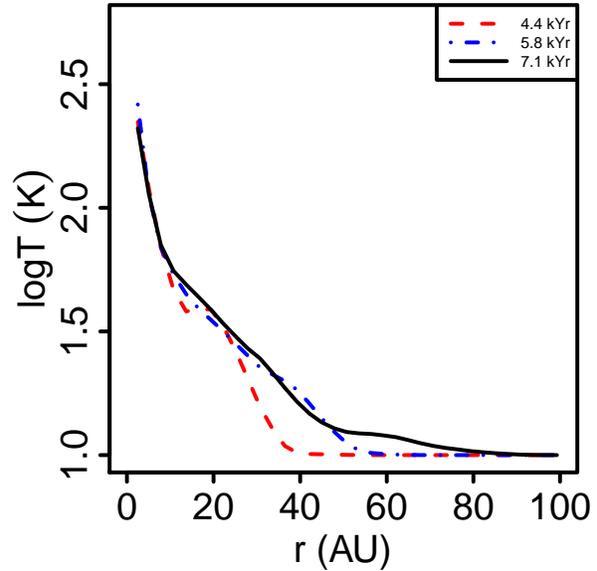}
\caption{ Azimuthally mass-averaged temperatures in the isolated system. At
later times the high $j$ gas has formed an extended disc which includes cold
interarm gas around the primary, leading to a break in the temperature profile.}
\label{fig:iso_temp1d}
\end{figure}

\begin{figure}
\includegraphics[width=\fw\textwidth]{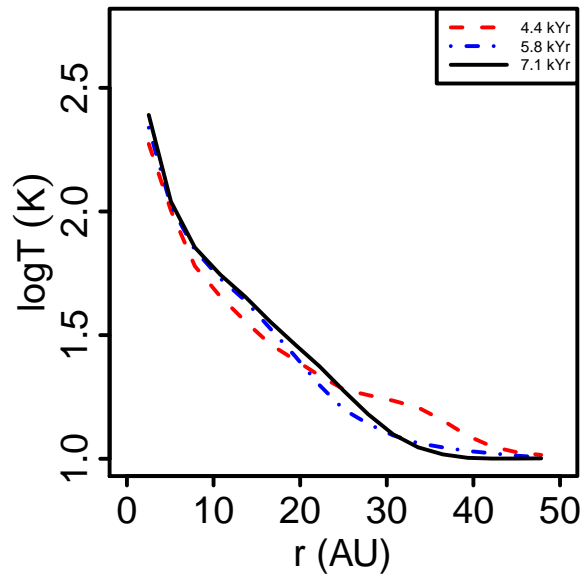}
\caption{ Azimuthally mass-averaged temperatures in the primary of the binary system. The disc
remains limited by tidal forces, and almost all bound gas is adiabatic. }
\label{fig:bin_temp1d}
\end{figure}

In Figs.~\ref{fig:iso_temp1d} \& \ref{fig:bin_temp1d} are plotted the
temperature profiles in the isolated and binary systems, respectively. At times
$\tpa$ and $\tpb$, the temperature declines steadily from $250\rm\,K$ to
$10\rm\,K$ at $40\rm\,AU$ and $\sim 60\rm\,AU$ respectively, whereupon it
remains at the temperature floor for larger radii, however there is little mass
at these radii (see Fig.~\ref{fig:miso}). At time $\tpc$, the disc becomes more
extended, and the temperature profile flattens at large radii.  The disk
contains a mixture of isothermal and adiabatic gas, with the spiral arms
containing most of the adiabatic component.  In the binary system
(Fig.~\ref{fig:bin_temp1d}), the
temperature profiles of the primary lack the low temperature, low density
component seen in the isolated system due to truncation of the disc at the tidal
radius, and do not evolve much after $\tpb$.

\subsubsection{Specific Angular Momentum}

\begin{figure}
\includegraphics[width=\fw\textwidth]{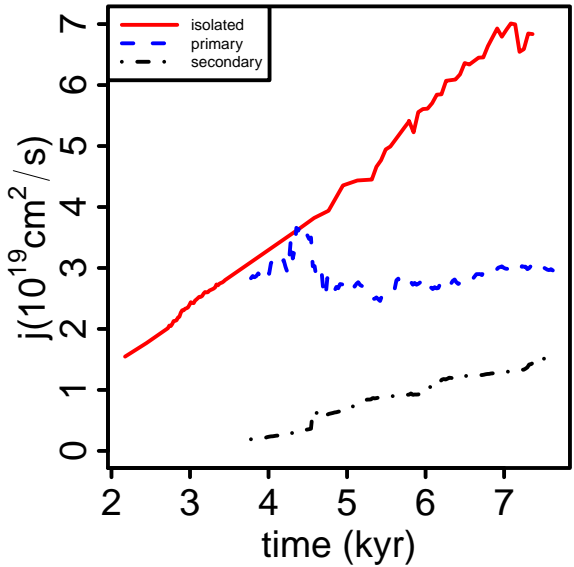}
\caption{ Evolution of the specific angular momentum $j$ of the isolated and binary systems.
          $j$ grows rapidly for the isolated system and leads to a more rapid buildup of an
          extended disc, whereas tidal effects and lower $j$ leads to smaller discs in the 
          binary system.}
\label{fig:jdisct}
\end{figure}

In Fig.~\ref{fig:jdisct} are plotted the combined disc and 
protostellar specific angular momenta for the isolated system, primary and
secondary. Each measure is computed in the centre-of-mass frame of the
given systems and is thus a measure of the spin (i.e. disc) component
of the specific angular momentum. The isolated system rapidly
increases in specific angular momentum,
and after $\tpc$, becomes unstable to fragmentation. In the binary system the
primary has a small spike in its specific angular momentum $j$ at $4.2\rm\,kyr$
due to weakly bound material appearing to be bound to the primary while the
secondary reaches pericenter. This is also seen in the mass plot at the same
time.  The primary exhibits very little evolution in $j$ thereafter, while from
Fig.~\ref{fig:mdisct} we see that it continues to accrete mass in the interim;
it accretes at constant $j$. In contrast the secondary increases steadily in $j$
along with $M$. 

Part of the reason that the binary increases more
slowly in specific angular momentum than the isolated system, is that
it is able to store momentum in the binary orbit. To measure this
effect,
we use the relationship
\begin{align} \label{eqn:j}
  {\bf L} = \sum_i {\bf R}_i {\bf\times} {\bf P}_i = {\bf R} {\bf\times} {\bf P}
  + \sum_i {\bf r}_i {\bf\times} {\bf p}_i
\end{align} 
where {\bf L} is the angular momentum, ${\bf R}_i$ are the positions
of each particle, ${\bf P}_i$ are the particle momenta, {\bf R} is the
barycentre, {\bf P} the velocity of the barycentre, ${\bf r}_i$ and
${\bf p}_i$ positions and momenta relative to the barycentric frame,
and decompose the binary into spin and centre of mass angular
momenta. We further get the specific angular momenta using the
appropriate mass-weightings along with Equation~\ref{eqn:j}. 
At $t = 7.1\rm\,kyr$, we find that in the barycentric frame,
the isolated system has a specific angular momentum of $7.0 \times
10^{19}\jmm$. In the binary system, the binary orbit itself has
$6.1 \times 10^{19}\jmm$, the primary has $3.0 \times 10^{19}\jmm$ and
the secondary $1.3 \times 10^{19}\jmm$ (also visible from
Fig.~\ref{fig:jdisct}). The specific angular momentum of the binary
orbit dominates that found in the other components.

We quantify alignment of the specific angular momentum of the disc with that of
the environment by computing the angle $\theta =
cos^{-1}(\bf{\hat{j}_d\cdot\hat{j}_e})$, with $\bf{\hat{j}_d}$ being the unit vector
of the specific angular momentum of the disc and $\bf{\hat{j}_e}$ being that of the
material in the environment in a box of dimensions $2000^3{\rm\,AU^3}$
surrounding the system. At the time of formation of the system
(Figs.~\ref{fig:lscale} \& \ref{fig:inset}) we find $\theta = 36^\circ$. In the
isolated system we observe angles of $13^\circ$, $7.7^\circ$, $12^\circ$ at
times $\tpa$, $\tpb$ and $\tpc$, respectively. In the binary system the
$\bf{\hat{j}_d}$ of the primary is $18^\circ$, $7.3^\circ$, $5.3^\circ$
at times $\tpa$, $\tpb$ and $\tpc$. The vectors tend to be oriented
perpendicular to the major axes of the filaments feeding the disc. Such
relatively small misalignments are consistent with other works suggesting filaments
tumbling about one of their shorter axes as the origin of protostellar angular
momentum \citep{walch09} and with observations of protostellar
systems on these length scales, which find that most gas is
extended in the direction perpendicular to outflows
\citep{tobinetal10}.

\subsection{Stability and Fragmentation}
\label{sec:stab}

\begin{figure}
\includegraphics[width=\fw\textwidth]{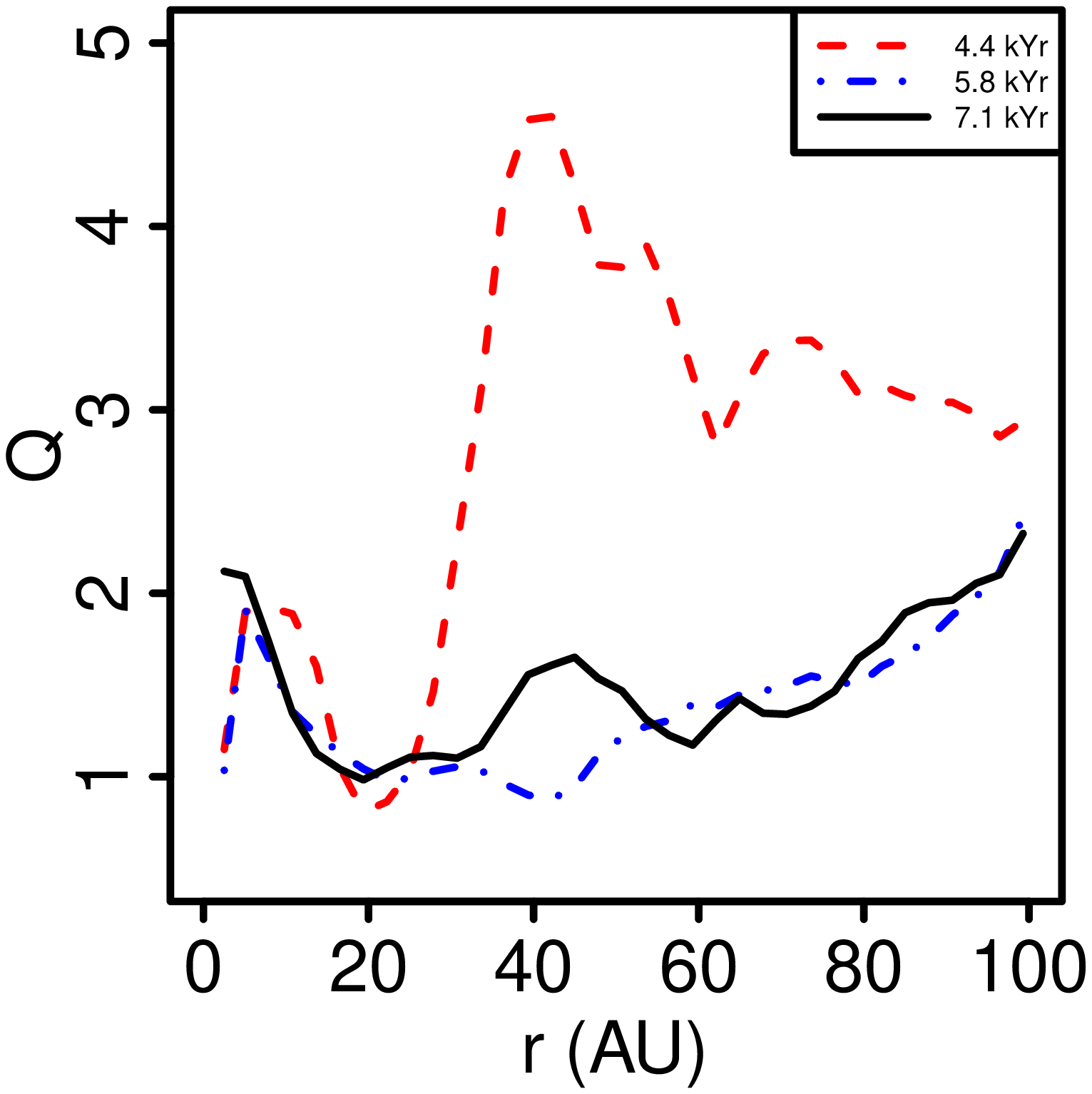}
\caption{ Azimuthally averaged Toomre Q for the isolated system at times 
          $\tpa$, $\tpb$ and $\tpc$. }
\label{fig:iso_q1d}
\end{figure}

\begin{figure}
\includegraphics[width=\fw\textwidth]{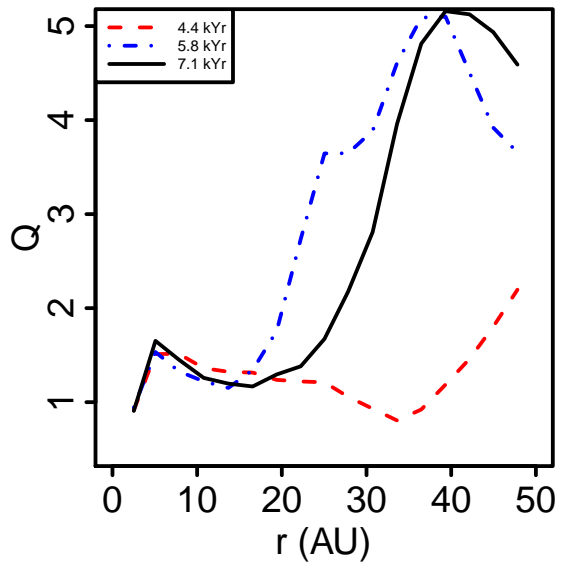}
\caption{ Azimuthally averaged Toomre Q for the binary system at times 
          $\tpa$, $\tpb$ and $\tpc$. The binary system remains stable for as 
          long as the simulation was run ($20\rm\,kyr$).}
\label{fig:binary_q1d}
\end{figure}

Throughout the simulation(s), the discs of both the isolated and binary systems
are massive and gravitationally unstable. To compare stability we compute the Toomre 
$Q = c_s \kappa/\pi G\sigma$ parameter  
locally using mass-weighted projections, through a $200 \times 200 \times 200\rm\,AU^3$ box, followed by azimuthal averages, of the
disc. The results are shown in Figs.~\ref{fig:iso_q1d} \&
\ref{fig:binary_q1d}. Although the $Q$ curve for the isolated system drops
below unity in some regions at $\tpa$ it does not fragment right away. As the
envelope is accreted, the disc evolves to a state where $Q$ lies between one and
two out to $90\rm\,AU$. The disk remains in the unstable regime, but $Q$ varies substantially as the system accretes.
In contrast, the
primary of the binary system has a minimum $Q$ of $1.3$ at later times and, except
initially, rises above two at radii greater than $25\rm\,AU$. The binary system
is thus significantly more stable than the isolated one, and this is borne out
by the fact that it does not fragment at later times (up to
$20\rm\,kyr$).

\begin{figure}
\includegraphics[width=\fw\textwidth]{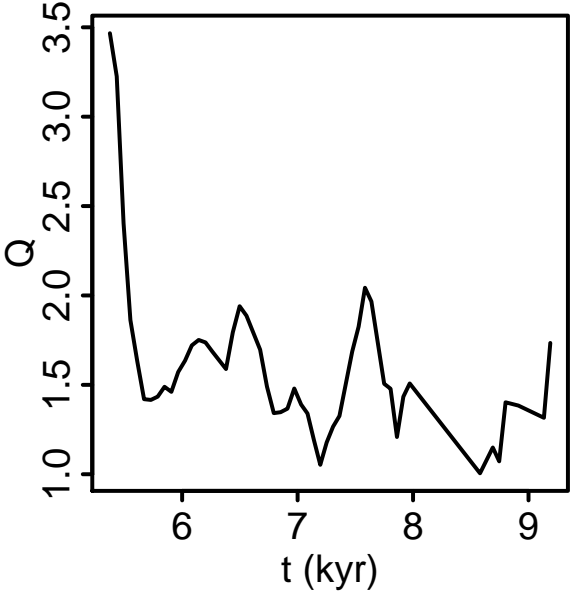}
\caption{ Toomre $Q$ paramater of the isolated system averaged in an annulus 
          from $60$--$70\rm\,AU$ in times preceeding clump formation. 
          A clump later forms at $\tpd$ at $65\rm\,AU$. }
\label{fig:iso_qt}
\end{figure}

\begin{figure}
\includegraphics[width=\fw\textwidth]{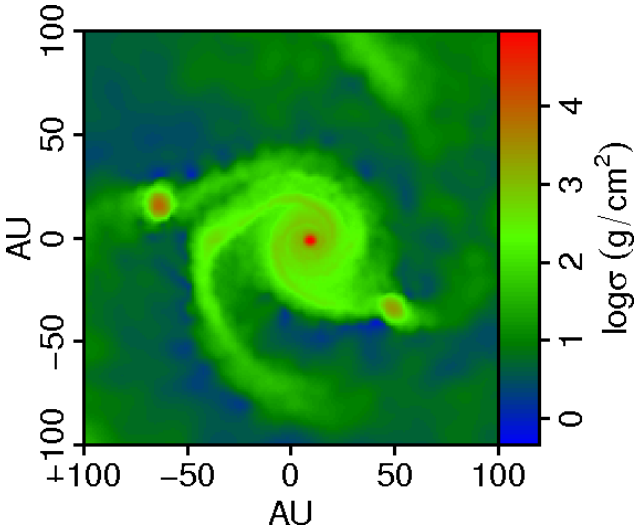}
\caption{ A surface density map of the isolated system at $t = \tpf$,
  approximately one orbital time after the second fragment formed. }
\label{fig:iso_frag}
\end{figure}

At $\tpd$ we find that the isolated system becomes unstable to fragmentation,
forming two clumps, one at $r = 65\rm\,AU$, followed by another at $r =
100\rm\,AU$ in the disc. The first clump (C1) formed in one of the material arms of
the disc from adiabatic gas, with an initial mass of $\mcla$ and specific
angular momentum of $1.8\times 10^{18}\jmm$.  The second clump (C2) formed at $\tpe$,
had a mass of
$\mclb$ and specific angular momentum of $3.2\times 10^{18}\jmm$. 
We seek to understand formation of the first
clump from adiabatic gas by looking at the behavior of $Q$ in the run-up to
fragmentation. In Fig.~\ref{fig:iso_qt} is plotted the $Q$ parameter averaged in
an annulus ranging from $60$--$70\rm\,AU$. $Q$ starts off initially very high
but drops as continuing accretion builds up the disc in this region.
Both clumps
are depicted in Fig.~\ref{fig:iso_frag} at $\tpf$. At this time, clump C1 has
made approximately 1.5 orbits and has a mass of $\mclaa$, and clump C2 one
half-orbit, has grown to $\mclbb$. Both clumps accrete at a rate of roughly 
$\mslyr{10^{-5}}$.

\section{\Rname{B}: Results}

\begin{figure*}
\includegraphics[width=\fw\textwidth]{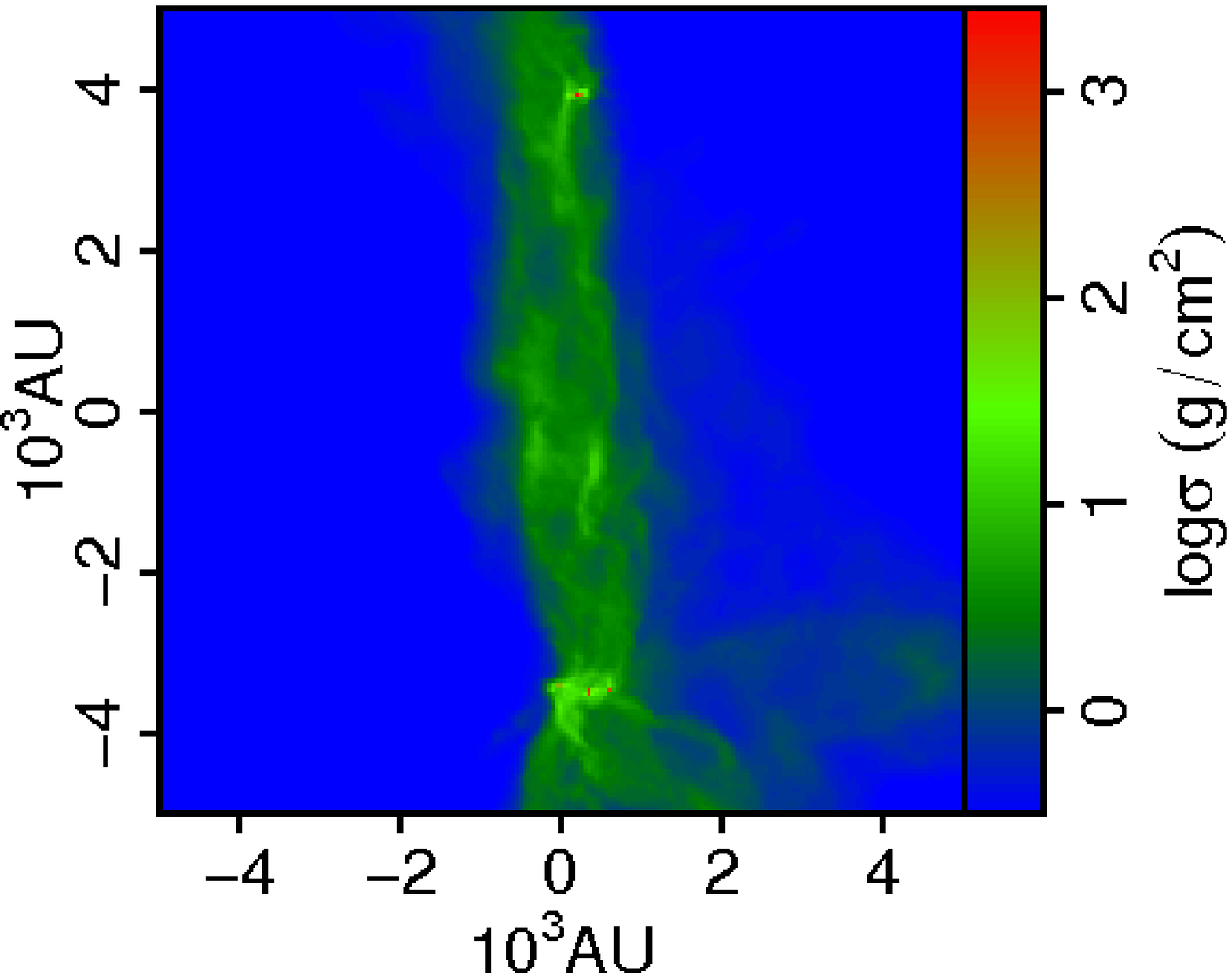} 
\includegraphics[width=\fw\textwidth]{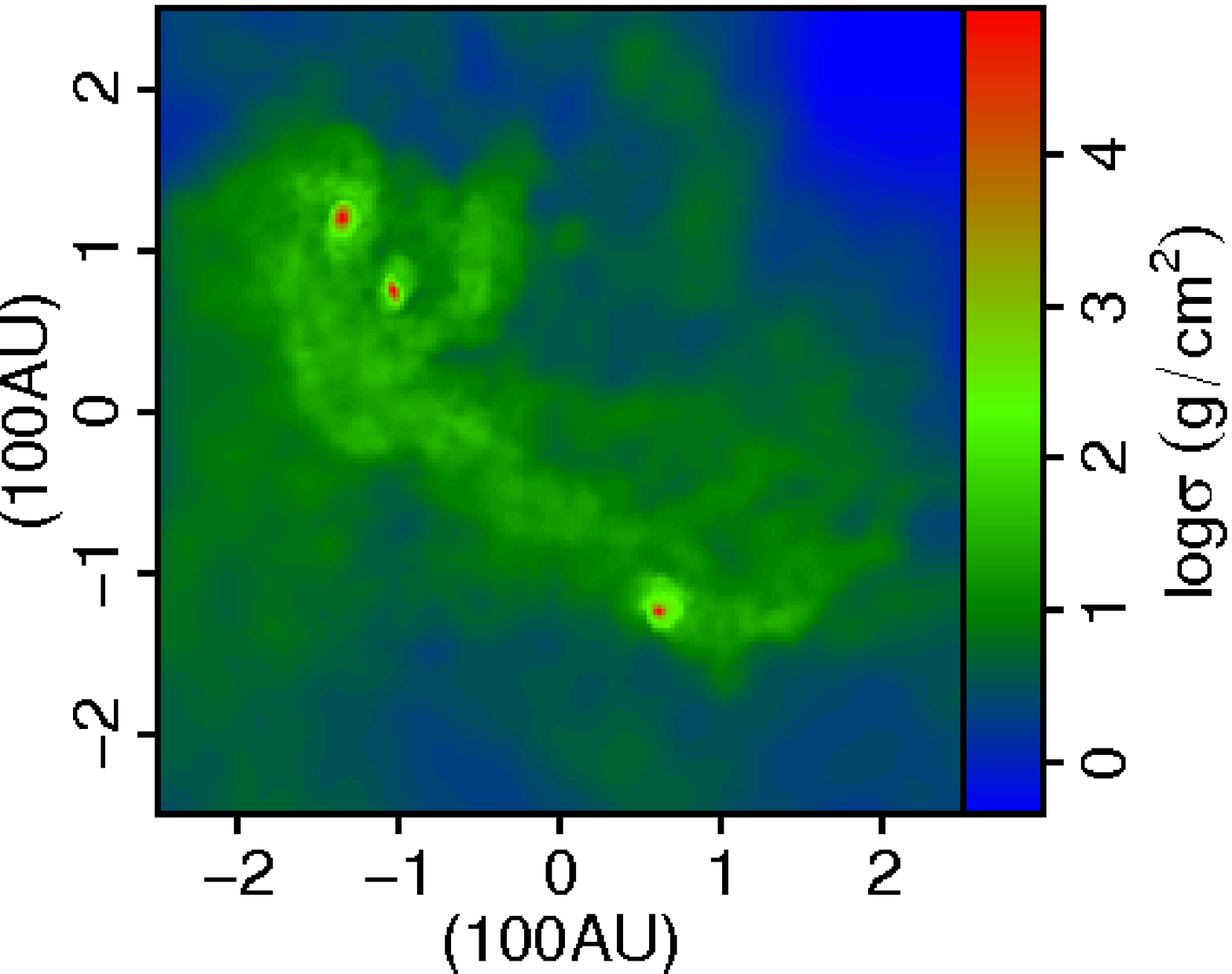}

\includegraphics[width=\fw\textwidth]{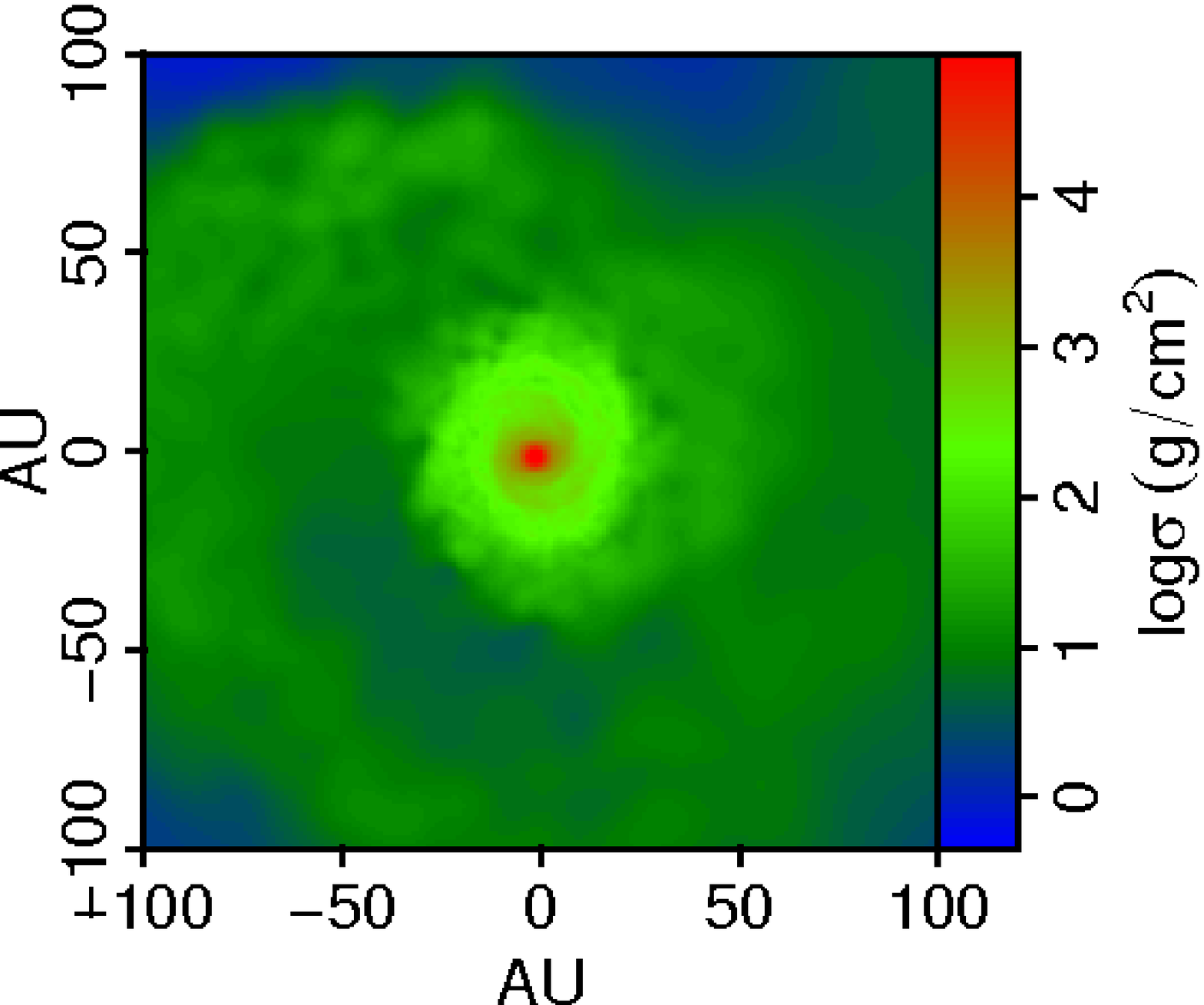} 
\includegraphics[width=\fw\textwidth]{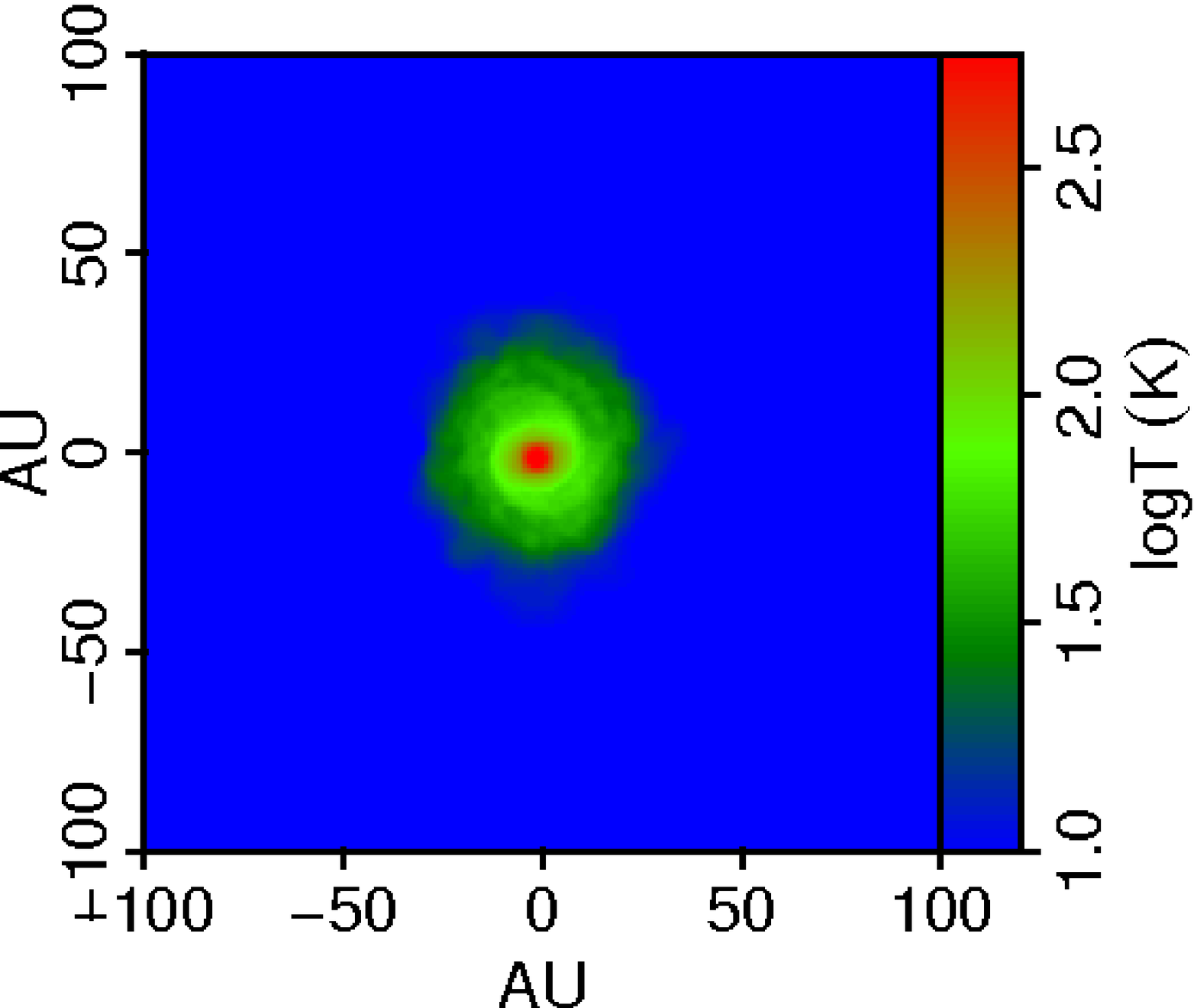}

\includegraphics[width=\fw\textwidth]{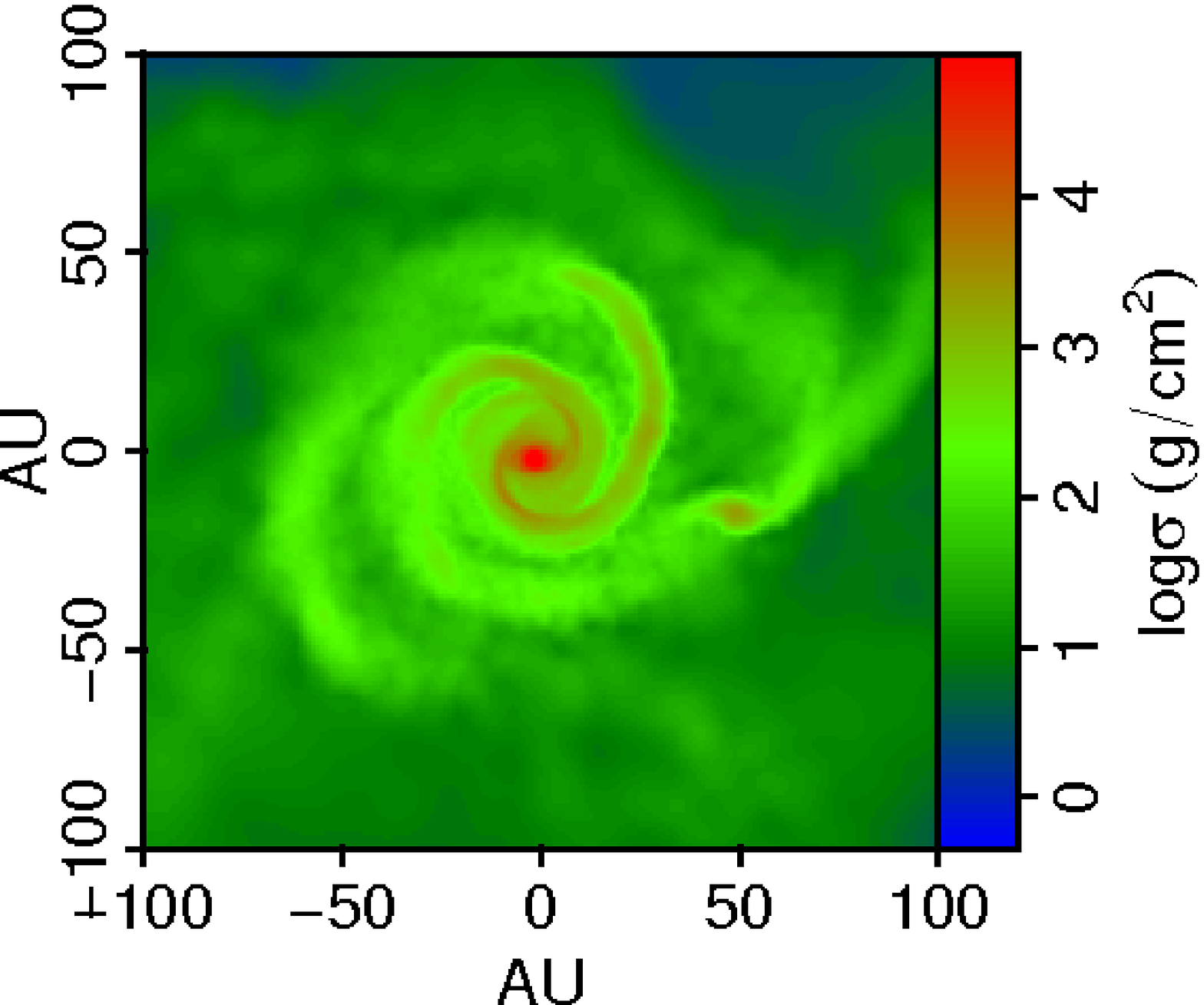} 
\includegraphics[width=\fw\textwidth]{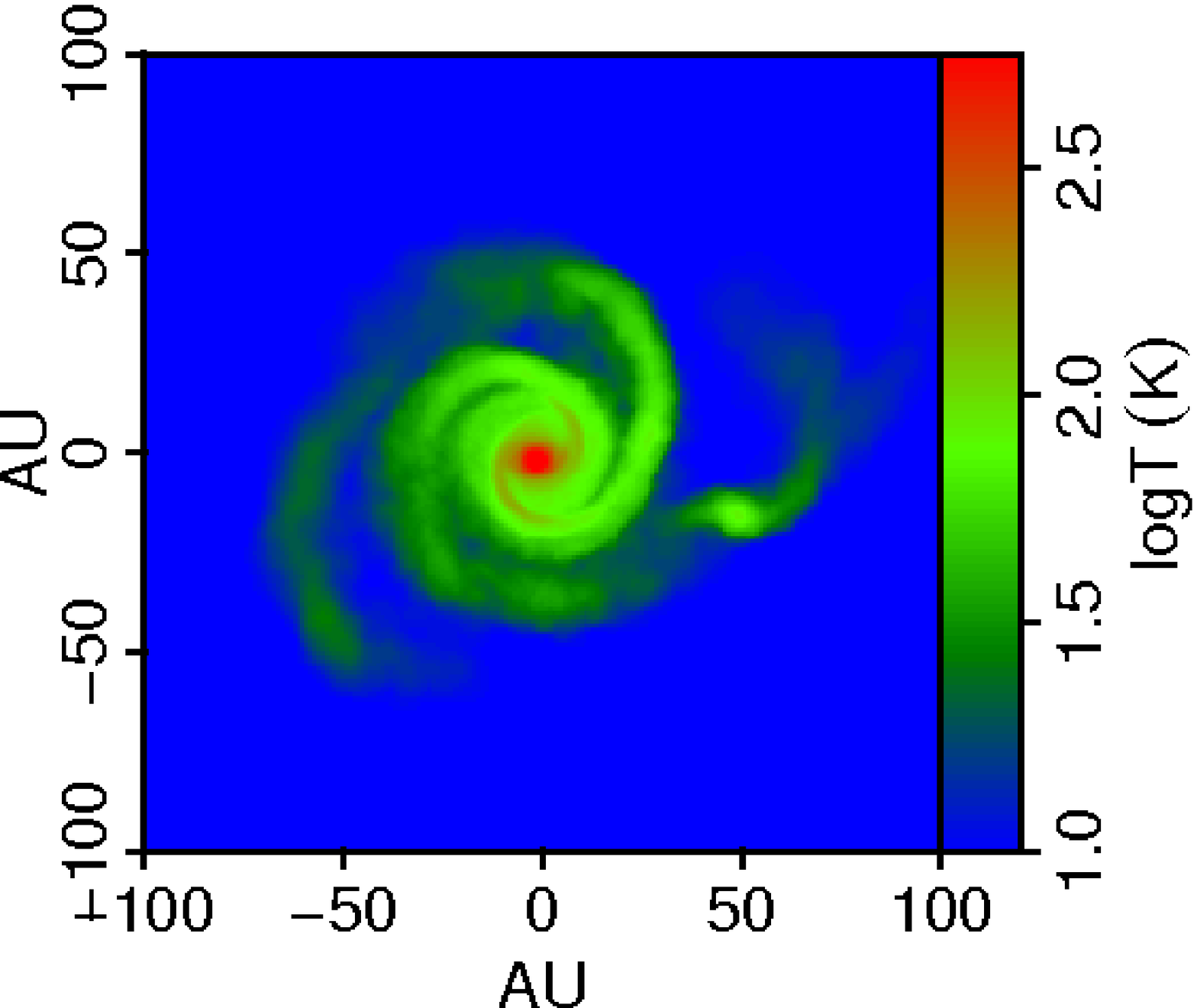}

\caption{ First row, Left: Surface density map of filament formed in \rname{B}.
  At the top and bottom of the filament are the isolated system and cluster, respectively.
  Right: The cluster as seen at the end of the simulation. One member, $~10^3\rm\,AU$ away, is not shown.
  Second and third rows:
  The surface densities and mass-weighted temperature maps of
  the isolated system, respectively in the left and right columns,
  shown at times $\alttpb$, and~$\alttpc$, after formation.}
\label{fig:alt_combipanel}
\end{figure*}

We describe here the results from a complimentary simulation, \rname{B}.
At approximately $t = \alttcol$ fragments form at either end of a long ($\altfillen$) 
filament, within $\sim \altfilfragdt$ of each other. In the following discussion and figures, we set
$t = \alttcol$ as being $t = 0$, as it is at this time when the first system, the isolated system, forms.

The upper left image in Fig.~\ref{fig:alt_combipanel} is a  
surface density plot of the filament, projected through a
$\altbox$ box. At the top end of the filament
lies the isolated system and the cluster is located at the bottom. The upper right image 
in Fig.~\ref{fig:alt_combipanel} is the surface density of the cluster at the end of simulation.
Only three of four members are shown. The two systems in the upper left corner of the image have
formed a binary, in the sense that they complete orbits around their centre of mass on a time scale
shorter than the cluster crossing time. For a more complete discussion of the cluster evolution,
see \ref{alt_clevo}.

In the second and third rows of Fig.~\ref{fig:alt_combipanel}, are the surface 
densities and mass-weighted temperature maps of the isolated system (lying at the top of the filament),
at $\alttpb$, and~$\alttpc$. In the isolated system we see the disc assemble rapidly from infalling 
filamentary gas, going through stages of increasing gravitational instability, until, as shown in the
bottom row of Fig.~\ref{fig:alt_combipanel}, gas trailing at the end of one of the material arms in the disc
fragments. As seen in the temperature maps, the gas in the fragment was in the adiabatic regime. The initial clump
mass is $\altmcl$.

\subsection{Cluster Evolution} \label{alt_clevo}
\begin{figure} \centering
\includegraphics[width=\fw\textwidth]{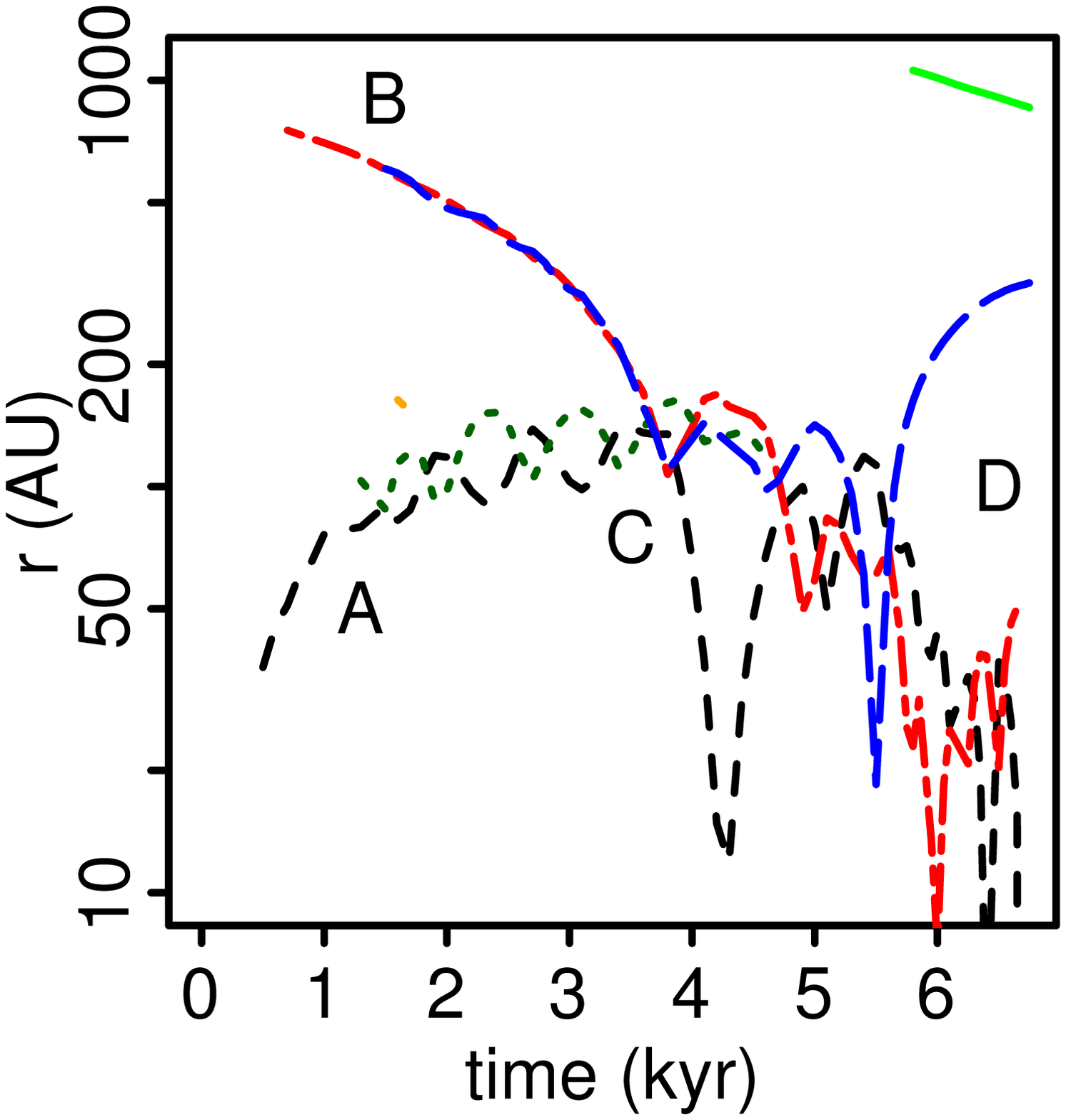}
\caption{ Distance from the origin of a fiducial cluster frame of reference vs. time, for various members
  of the cluster. The frame of reference is based on the trajectory of the oldest member, hence the
  separation approaches zero at the beginning and end of the data. Several objects pair up to form
  short lived 'binaries' while in the cluster. 
Two such binaries can be seen at points A and B in Fig.~\ref{alt_clevo},
through the mutual oscillation of their distances in the cluster frame. 
At point C, all extant members of the cluster briefly converge. Between C and D,
a merger occurs, leading to the formation of a new binary and the ejection of the third member.}
\label{fig:alt_clevo}
\end{figure}

In Fig.~\ref{alt_clevo} are plotted the distances of cluster members to the origin of a fiducial cluster frame vs. time.
Since the cluster is a rapidly evolving object at this stage, we choose the fiducial frame as being the one which
moves in a straight line from the original position of the first clump to its final position. We can see in 
Fig.~\ref{alt_clevo} that members are typically $\sim 100\rm\,AU$ from the origin in this frame. Members spend
most of their time in binary associations, whereby it is meant that they orbit around their common center of mass
on timescales shorter than the cluster crossing time. Two such binaries can be seen at points A and B in Fig.~\ref{alt_clevo},
through the mutual oscillation of their distances in the cluster frame. 
At point C in the figure, the members of the cluster converge briefly and their
 mutual separations become comparable to their binary separations. Between C and D, a merger occurs, and a new binary forms,
with the remaining member being sent on a larger orbit.

\subsection{Disc Evolution} \label{alt_devo}
\subsubsection{Disc Mass}

\begin{figure} \centering
\includegraphics[width=\fw\textwidth]{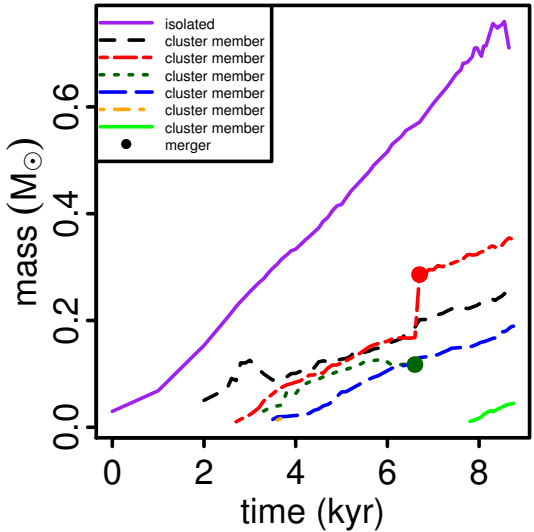}
\caption{ Total mass accreted (disc + protostar) versus time for the isolated system, 
and the cluster. A merger occurs between two prestellar systems in the cluster at $\altmergerdt$.}
\label{fig:alt_mdisct}
\end{figure}

In Fig.~\ref{fig:alt_mdisct}, the accretion histories of the isolated system, and 
those of all objects formed in the cluster are plotted. The isolated system accretes gas at a
high rate, at an average of $\mslyr{8 \times 10^{-5}}$. The accretion rates observed are quite
comparable to those for the isolated system from \rname{A}, and thus they are well above the
expected asymptotic rates ($m_\circ \gtrsim 10$).

For the cluster, the average accretion rate is $\sim \mslyr{3 \times 10^{-5}}$, with little
deviation between objects. The evolution of the cluster is complex. The pattern of fragmentation  
of core gas
never yielded objects which were isolated for very long, quickly leading to the production of 
a couple of short-lived binaries with similar evolution (similar accretion rates and 
generally limited disc size). A major merger occurs at $\altmergerdt$; the result of a 
head-on collision between two clumps. Given the chaos of the cluster, the accretion rates as
seen by the slopes in Fig.~\ref{fig:alt_mdisct} are remarkably uniform.

\subsubsection{Specific Angular Momentum}

\begin{figure}
\includegraphics[width=\fw\textwidth]{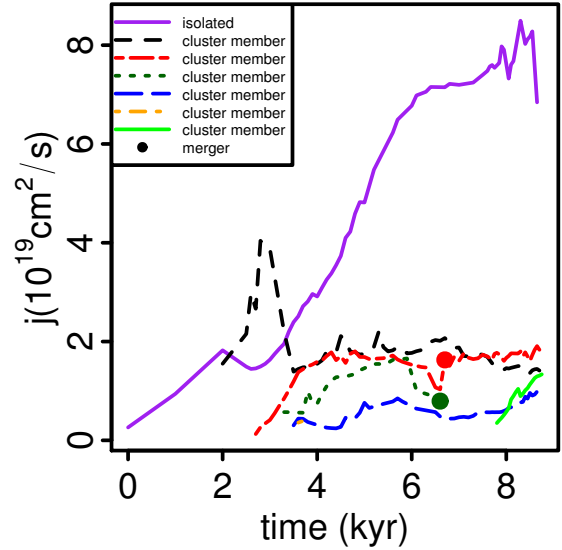}
\caption{ Evolution of the specific angular momentum $j$ of the isolated system and the cluster.
          $j$ grows rapidly for the isolated system and leads to a more rapid buildup of an
          extended disc, whereas various tidal effects and close encounters lower $j$ and leads to smaller discs in the 
          binary system.}
\label{fig:alt_jdisct}
\end{figure}

In Fig.~\ref{fig:alt_jdisct} are plotted the combined disc and 
protostellar specific angular momenta for the isolated system,  and for
the objects in the cluster. As in \rname{A}, the isolated system rapidly
increases in specific angular momentum. Upon reaching $\sim 7 \times 10^{19}\jmm$,
material arms form in the isolated system, moving angular momentum out of the system,
and allowing angular momentum to remain constant while accretion continues (compare to
Fig.~\ref{fig:alt_mdisct}). As the arms become stronger, the determination of $j$ 
becomes more difficult, because gas is ejected with increasing violence from the ends
of the material arms back into the envelope. About $1\rm\,kyr$ after the onset 
of this behaviour, the arm fragments (shown in Fig.~\ref{fig:alt_combipanel}).

The evolution of the cluster angular momenta shows a rather remarkable contrast. 
Following Fig.~\ref{fig:alt_jdisct} once more, we see that the first object to form in the 
cluster evolves as one might expect, with both mass and specific angular momentum increasing.
Once other fragments form in the cluster, the specific angular momentum of the system drops 
until hitting a maximum value which appears to characterise the entire cluster.  
The pattern of fragmentation in the core never left any object in the cluster truly isolated for very long,
with some short-lived binaries forming, sometimes ending in mergers, but generally lasting long
enough to tidally limit and alter the structure of the discs of cluster members.
Referring once more
to the mass evolution plots in Fig.\ref{fig:alt_mdisct}, we see that accretion continues, somewhat
abated by the competition amongst cluster members for infalling gas. We see then a stronger,
clearer, reiteration of the result from \rname{A}, namely that upon hitting a threshold,
accretion continues but at constant specific angular momentum.

The major merger occuring at $\altmergerdt$, indicated by the large
filled circles in Fig.~\ref{fig:alt_jdisct} has an interesting effect on the angular momenta
of the merging systems. Their momenta both dip quite suddenly at the time of collision. This 
is due to the ejection of plumes of high-$j$ gas at the moment of merger. The gas is then quickly accreted onto the 
new system, bringing it back up to the angular momentum ceiling for the cluster. 

\section{Discussion}
\subsection{Fragmentation}
\subsubsection{Accretion and specific angular momentum}
In the simulations performed the isolated systems become prone to fragmentation
via the development of massive material arms, even though the gas is
adiabatic. Instabilities in the isolated systems are driven by steady accretion
of gas with increasing specific angular momentum. The disc, sufficiently massive
to generate spiral arms, accumulates a reservoir of gas lying at large radii
(i.e. with high specific angular momentum). This reservoir is then swept up by
the spiral arms, which when sufficiently massive, become
prone to fragmentation; such instabilities have been documented in
similar contexts in the literature  (e.g. \citealt{bonnell94};
\citealt{whitworthetalrot95}; \citealt{hennebelleetalcomp04}). At the same time,
\citet{offnerklein08} observe that in simulations of freely decaying turbulence
the resulting protostellar systems are more prone to fragmentation, and
attribute it to high rates of accretion. Seeing as the isolated systems in the
present study have accretion rates several times that of the nominal 1D asymptotic
rate, perhaps the tendency toward
fragmentation observed by \citet{offnerklein08} is due to a combination of
higher accretion rates {\em and} accretion of gas with high specific
momentum. Furthermore, it could be that the continued driving of turbulence is reducing
the specific angular momentum of accreted gas, or acting as a viscosity and depleting the
reservoirs of gas which would otherwise be swept up in dense material arms.

The binary system from \rname{A} and the systems in the cluster from \rname{B} remain
 stable throughout the simulation. Although the discs
are sufficiently massive to be self-gravitating, and the systems are accreting, a
number of effects work against outer disc fragmentation. The tidal limiting
of the discs, as discussed in section \ref{sec:cumass}, is a major barrier to
fragmentation. In \rname{A}, although the binary separation evolves (increases)  throughout the
simulation, by the end of our run the separation is $\sim 75$ AU, and given the
tendency of the binary mass ratio to approach unity
(e.g. \citealt{batebonnell97}), there is very little chance for outer disc
fragmentation to occur. In \rname{B} the cluster appeared to have an effective ceiling
in the specific angular momentum for member systems.
Longer simulations looking at the long term evolution of multiple systems and clusters
 should be undertaken to see if they separate
sufficiently to allow extended discs to grow. Even if at some later point they do grow, 
fragmentation is less likely
because most mass will likely have made it onto the star and a high disc/star mass ratio will be harder
to achieve, also, accretion rates decrease as time progresses.
In \rname{A}, the tendency for the secondary to accrete high specific angular momentum gas as
seen in this paper and in other studies of binaries
(e.g. \citealt{batebonnell97}), acts strongly to deplete the reservoir of gas in
the outer part of the system and shuts out the possibility for the growth of the
same Toomre instability which lead to fragmentation of the isolated system.
Furthermore, the enhanced inward transport of mass due to tidal torques shown in
\ref{sec:cumass} acts to stabilise the disc against fragmentation, an effect
that has also been observed by (e.g. \citealt{mwqs05}).

\subsubsection{Accretion and gas temperatures}
In the following we discuss analyse \rname{A} in more detail.
In the isolated system two clumps form by the end of the simulation. 
They form in the outer reaches of the disc where temperatures were
low, and where some interarm gas is still isothermal at 10\,K, although clump C1
 forms in a material arm from adiabatic gas with a
temperature of $\sim 30\rm\,K$. Recent simulations of star formation including a
flux-limited diffusion treatment for radiative transfer and some modelling of
protostellar accretion emphasise the importance of radiative feedback on the 
envelope. To estimate the effect we compute the accretion luminosity of
the protostar \citep{stahlerpalla05}: 
\begin{align} \label{eqn:lacc}
  L_{acc} &= \frac{GM_{\star}\dot{M}}{R_{\star}} \notag\\
          &= {61\rm\,L_\odot}
  \left(\frac{\dot{M}}{10^{-5}\rm\,M_\odot\,yr^{-1}}\right)
  \left(\frac{M_\star}{\msl{1}}\right)
  \left(\frac{R_\star}{5\rm\,R_\odot}\right)^{-1}
\end{align} 
with $M_\star = \msl{0.11}$ being the protostellar mass, which we
compute by taking all mass within $r <= 5{\rm\,AU} \simeq 2\epsilon$;
$\dot{M} = \mslyr{5\times 10^{-6}}$ is the (time-averaged) accretion
rate onto the protostar, and $R_\star$ the protostellar radius that,
since it is sub-grid, we assume a fiducial value of $5 R_\odot$. From
this we get a typical accretion luminosity of $L_{acc} =
3.4\rm\,L_\odot$. Clump C1 forms at $r_c = 65\rm\,AU$. To
try to get an upper limit on the temperature at that radius, we
assume that the disc presents an absorbing surface of height $h =
10\rm\,AU$ spanning $r_c \pm 5\rm\,AU$, and that all of the
accretion luminosity crossing this surface deposits all of
its energy in this region, yielding a heating rate of $L_{acc} h/2r_c =
10^{33}\rm\,erg\,s^{-1}$. We assume \citet{dalessiocalvethartmann01}
opacities and a $1\rm\,\mu m$ grain size, and a background
irradiation temperature of $30\rm\,K$, and using the existing
density field of the gas within $r_c \pm 5\rm\,AU$ at the time of
fragmentation, we calculate the temperature at which the cooling rate
balances the heating rate of the gas to be $50\rm\,K$. We wish to
emphasise that the assumptions made here are extremely conservative
with the aim of bracketing the upper end of conceivable outer disc
temperatures. Given the weak
dependence of $Q$ on $T$, it does not appear that the accretion
luminosity will be able to affect the disc stability at these large
radii.

\subsubsection{Clump masses and accretion rates}
Again, we limit the discussion here to \rname {A} for brevity.
The isolated system forms two clumps of masses $\mcla$ and
$\mclb$. What kind of clump masses should one expect in the outer
disc? Because the disc is in a disordered state at the time of clump
formation we do not compare the clump masses to those computed from
axisymmetric models, but instead we compute the local Jeans mass of
the clump-forming gas, which for clump C1 was
$4.7\rm\,M_{jup}$, and that of clump C2 was $4.5\rm\,M_{jup}$.
The particle mass in the simulations was $\msl{10^{-5}}$, hence in the
clumps the Jeans mass is resolved with $\sim 450$ particles, or $\sim
14$ smoothing kernels. The fragmentation phenomena are thus
well-resolved according to the criteria set forth in the literature
(\citealt{batejm97}; \citealt{nelson06}).  
 In addition, the
initial angular momentum radii of the clumps are 0.31\,AU and 0.73\,AU,
respectively, and should they undergo a second collapse in $\sim 10^4$
years due to $\rm H_2$ dissociation, any accretion onto the final body
will take place through an accretion disc \citep{boleyetal10}. We
observe final clump masses of $\mclaa$ and $\mclbb$ for C1 and C2,
respectively, however whether or not dissociative collapse will occur
before these masses are achieved is highly sensitive to assumptions
about the dust.
We emphasise here that the connection between our final obtained clumps 
and the substellar companions (the planets or brown dwarfs) produced from them is 
uncertain. A dissociating collapse of the clumps and the establishment of an
accretion disc is a certainty, however other dynamical instabilities may form, such as a bar
instability, and can play a role in redistributing mass and angular
momentum. Because the simulation lacks the necessary physics and resolution to
follow the internal evolution of the clumps, we refrain from speculating on the
final object masses.
Nevertheless, we can estimate accretion luminosities for our final clumps
using our clump
parameters. 
Both clumps C1 and C2 had accretion rates of $\sim
\mslyr{10^{-5}}$. Computing the accretion luminosity using
Equation~\ref{eqn:lacc}, inputting the measured accretion rate,
clump masses of 5.5--39$\rm\,M_\odot$, and assuming a fiducial final
size of ${1\rm\,R_{jup}} = 
7\times10^{9}\rm\,cm$, gives accretion luminosities ranging between
16--110$\rm\,L_\odot$. Such relatively high luminosities are achieved
at this accretion rate because the radius of Jupiter is $\sim 50$
times smaller than our fiducial one for the central protostar
($5\rm\,R_\odot$). With such extremely high luminosities we can see that mass
accretion simply cannot
continue unabated without feedback playing some role. Our final clump masses
and luminosities thus fall on the upper end of the range of expected values.

\section{Conclusions}
The primary motivation of this paper was to compare the evolution of
an isolated and binary/multiple systems during the early stages of prestellar
development within the same environment. Of particular interest was
the potential for fragmentation of such early, accreting and self-gravitating
discs. The following are our findings:

\begin{itemize}
\item The initial collapse occurs within filaments that tend to collimate infalling gas onto 
      the central prestellar objects, whose specific angular momenta  tend to remain
      aligned to that of the environment (and perpendicular to the embedding filament)
      on scales of $\sim 1000\rm\,AU$.
\item During formation, isolated systems must accrete
  mass and increase their specific angular momentum, leading to the formation of massive, extended
  discs, which undergo strong gravitational instabilities and are susceptible to fragmentation
\item In systems starting out as binaries and multiples, an effective specific angular momentum 
  ceiling exists, limiting the maximum $j$ of the systems but not their mass, making disc fragmentation 
  unlikely. The ceiling is the product of tidal interactions which strip cold gas from members, and concentrate their
  mass profiles, and the redirection of angular momentum into orbits.
\end{itemize}

\section{Acknowledgements}
The computations were performed in part on the Brutus cluster at the ETH Z\"urich,
and the zBox2 and zBox3 supercomputers
at the ITP, University of Z\"urich.

\bibliographystyle{mn2e}
\bibliography{isobin}

\end{document}